\documentclass[9pt,twocolumn]{extarticle}
\pdfoutput=1

\usepackage{soul}
\usepackage{titlesec} 
\usepackage[backend=bibtex,style=science]{biblatex}
\DeclareNameAlias{sortname}{first-last}
\DeclareNameAlias{default}{first-last}
\bibliography{biblio} 
\usepackage{dsfont}
\usepackage{graphicx}
\usepackage{subcaption}
\usepackage[margin=0.9in]{geometry}
\usepackage[usenames,dvipsnames]{color}
\usepackage[colorlinks,linkcolor=Blue,urlcolor=Blue,citecolor=Blue]{hyperref}
\usepackage{amsmath,amssymb}
\usepackage{amsfonts}
\usepackage{bbm}
\usepackage[squaren]{SIunits}

\usepackage{mathpazo} 
\usepackage{courier} 
\normalfont

\usepackage[T1]{fontenc}
\usepackage{caption}
\renewcommand{\thefigure}{\arabic{figure}}

\usepackage{upgreek}

\newcommand{\ket}[1]{\ensuremath{|#1\rangle\mkern-1mu}}
\newcommand{\bra}[1]{\ensuremath{\mkern-1mu\langle#1|}}

\newcommand{\Tr}[1]{\mathrm{Tr}#1}
\newcommand{\proj}[1]{\ket{#1}\!\bra{#1}}

\newcommand{\ad}[1]{\textsuperscript{#1}\kern-2pt}

\usepackage{amsmath,amsthm,mathtools}
\newtheorem{theorem}{Theorem}
\DeclarePairedDelimiter{\set}{\lbrace}{\rbrace}
\newcommand{\C}{\mathbb{C}}
\renewcommand{\vec}[1]{\mathbf{#1}}

\def\({\left(}
\def\){\right)}
\def\[{\left[}
\def\]{\right]}

\makeatletter
\def\blx@maxline{77}
\makeatother

\usepackage{capt-of}

\usepackage[ruled]{algorithm2e} 
\usepackage{newfloat,algcompatible} 
\def\({\left(}
\def\){\right)}
\def\[{\left[}
\def\]{\right]}

\def\mytitle{ 
Multidimensional quantum entanglement with large-scale integrated optics 
\vspace{-4mm}}      
 
\title{\vspace{-1.5cm}\huge\textbf{\textrm{\mytitle}}}  
 
\date{} 
\begin{document}
\twocolumn[{
\maketitle 
\vspace{-12mm}

\begin{center}
\begin{minipage}{1\textwidth}
\begin{center}
\author{\large{Jianwei Wang$^{1,2\dagger}$, Stefano Paesani$^{1\dagger}$, Yunhong Ding$^{3,4\dagger}$, Raffaele Santagati$^{1}$, Paul Skrzypczyk$^{5}$, Alexia Salavrakos$^{6}$, Jordi Tura$^{7}$, Remigiusz Augusiak$^{8}$, Laura Man\v{c}inska$^{9}$, Davide Bacco$^{3.4}$, Damien Bonneau$^{1}$, Joshua W. Silverstone$^{1}$, Qihuang Gong$^{2}$, Antonio Ac\'in$^{6,10}$, Karsten Rottwitt$^{3,4}$, Leif K. Oxenl{\o}we$^{3,4}$, Jeremy L. O\textquoteright Brien$^{1}$, Anthony Laing$^{1}$, Mark G. Thompson$^{1}$}} 
\end{center}
\end{minipage}
\end{center}

\vspace{0mm}

\begin{center}
\begin{minipage}{0.95\textwidth}
\begin{center}
\textit{\textrm{
\textsuperscript{1} 
Quantum Engineering Technology Labs, H. H. Wills Physics Laboratory and Department of Electrical and Electronic Engineering, University of Bristol, BS8 1FD, Bristol, United Kingdom
\\\textsuperscript{2} State Key Laboratory for Mesoscopic Physics, School of Physics, Collaborative Innovation Center of Quantum Matter, Peking University, Beijing 100871, China
\\\textsuperscript{3} Department of Photonics Engineering, Technical University of Denmark, 2800 Kgs. Lyngby, Denmark
\\\textsuperscript{4} Center for Silicon Photonics for Optical Communication (SPOC), Technical University of Denmark, 2800 Kgs. Lyngby, Denmark
\\\textsuperscript{5} H. H. Wills Physics Laboratory, University of Bristol, BS8 1TL, Bristol, United Kingdom
\\\textsuperscript{6} ICFO-Institut de Ciencies Fotoniques, The Barcelona Institute of Science and Technology, 08860 Castelldefels, Barcelona, Spain
\\\textsuperscript{7} Max-Planck-Institut f\"ur Quantenoptik, Hans-Kopfermann-Stra{\ss}e 1, 85748 Garching, Germany
\\\textsuperscript{8}
Center for Theoretical Physics, Polish Academy of Sciences, Aleja Lotnik\'ow 32/46, 02-668 Warsaw, Poland
\\\textsuperscript{9} QMATH, Department of Mathematical Sciences, University of Copenhagen, Universitetsparken 5, 2100 Copenhagen {\O}, Denmark
\\\textsuperscript{10} ICREA - Instituci\'{o} Catalana de Recerca i Estudis Avan\c{c}ats, Pg. Llu\'{i}s Companys 23 Barcelona, 08010, Spain
\\\textsuperscript{$\dagger$} These authors contributed equally to this work. \\
Emails: 
jianwei.wang@bristol.ac.uk; 
yudin@fotonik.dtu.dk; 
anthony.laing@bristol.ac.uk; mark.thompson@bristol.ac.uk 
\\
}}
\end{center}
\end{minipage}
\end{center}
 
\setlength\parindent{12pt}
\begin{quotation}
\noindent 
{
The ability to control multidimensional quantum systems is key for the investigation of fundamental science and for the development of advanced quantum technologies. 
Here we demonstrate a multidimensional integrated quantum photonic platform able to robustly generate, control and analyze high-dimensional entanglement. 
We realize a programmable bipartite entangled system with dimension up to $15 \times 15$ on a large-scale silicon-photonics quantum circuit. The device integrates more than $550$ photonic components on a single chip, including $16$ identical photon-pair sources. We verify the high precision, generality and controllability of our multidimensional technology, and further exploit these abilities to demonstrate key quantum applications experimentally unexplored before, such as quantum randomness expansion and self-testing on multidimensional states. Our work provides a prominent experimental platform for the development of multidimensional quantum technologies.
}
\end{quotation}}]

As a generalization of two-level quantum systems (\textit{qubits}),  multidimensional quantum systems (\textit{qudits}) exhibit distinct quantum properties and can offer significant improvements in key applications. 
For example, qudit systems allow higher capacity and noise robustness in quantum communications~\cite{PhysRevLett.88.127902, 
OAM:Tele,Boucharde1601915, Islame1701491}, can be used to strengthen the violations of generalized Bell and Einstein-Podolsky-Rosen (EPR) steering inequalities~\cite{PRL_Doherty,CGLMP, SATWAP}, provide richer resources for quantum simulation~\cite{Spin-1Simulation, Superconducingqudit}, and offer higher efficiency and flexibility in quantum computing~\cite{Lanyon:2008gv,QuditQC, QuditQCfactoring}. Moreover, encoding and processing qudits can represent a more viable route to larger Hilbert spaces. 
These advantages motivate the development of multidimensional quantum technologies in a variety of systems, such as photons~\cite{Mair:2001,Dada:2011}, superconductors~\cite{Superconducingqudit,Svetitsky2014}, and atomic systems~\cite{IonsQudit, NVQudit}. 
While complex interaction engineering and control sequences are required to encode and manipulate superconducting and atomic qudits, photons represent a promising platform able to naturally encode and process qudits in various degrees of freedom, e.g., orbital angular momentum (OAM)~\cite{Mair:2001, Erhard2017}, temporal modes~\cite{Gisin:timebin, Islame1701491}, and frequency~\cite{Xie:frequency,Roberto:frequency}. 
Previous pioneering work on qudits include realizations of complex entanglement~\cite{Malik:2016}, entanglement in ultra-high dimension~\cite{Fickler640}, and practical applications in quantum communication~\cite{Islame1701491,OAM:Tele,Boucharde1601915} and computing~\cite{Spin-1Simulation, Superconducingqudit,Lanyon:2008gv}.  
However, these approaches present limitations in terms of controllability, precision and universality, which represent bottlenecks for further developments of multidimensional technologies. 
For example, the arbitrary generation of high-dimensional entanglement is a key experimental challenge, typically relying on complex bulk-optical networks and post-selection schemes~\cite{Dada:2011,Roberto:frequency,Gisin:timebin,Xie:frequency,Erhard2017}.  
In general, these approaches lack the ability to perform arbitrary multidimensional unitary operations with high fidelity~\cite{Erhard2017, Malik:2016}, a key factor in  quantum information tasks. 
Integrated microring resonators able to emit multidimensional OAM ~\cite{IOAM} and frequency ~\cite{Roberto:frequency} states have been reported, but these present limited fidelity and difficulties for on-chip state control and analysis, thus not fully exploiting the high precision, scalability and programmability of integrated optics. 

Here we report a multidimensional integrated quantum photonic device that is able to generate, manipulate and measure multidimensional entanglement fully on-chip with unprecedented precision, controllability and universality. 
Path-encoded qudits are obtained having each photon exist over $d$ spatial modes  simultaneously, and entanglement is produced by a coherent and controllable excitation of an array of $d$ identical photon-pair sources. 
This allows the generation of multidimensional entangled states with an arbitrary degree of entanglement. 
Universal operations on path-encoded qudits are possible in linear-optics for any dimension ~\cite{Reck, Oxford:U,Carolan:15}, and our device performs arbitrary multidimensional projective measurements with high fidelity. 
  
By integrating more than 550 components on a chip, which embeds 16 identical photon sources and 93 programmable phase controls, 
we demonstrate the generation, manipulation and measurement of entangled states of two-photon with dimension up to $15\times 15$. 
The capabilities achieved allow us to demonstrate high-quality multidimensional quantum correlations (verified by Bell non-locality and EPR steering) and to implement experimentally unexplored multidimensional quantum information protocols: multidimensional randomness expansion and state self-testing.

\begin{figure*}[ht!] 
\centering
\includegraphics[width=1.\textwidth]{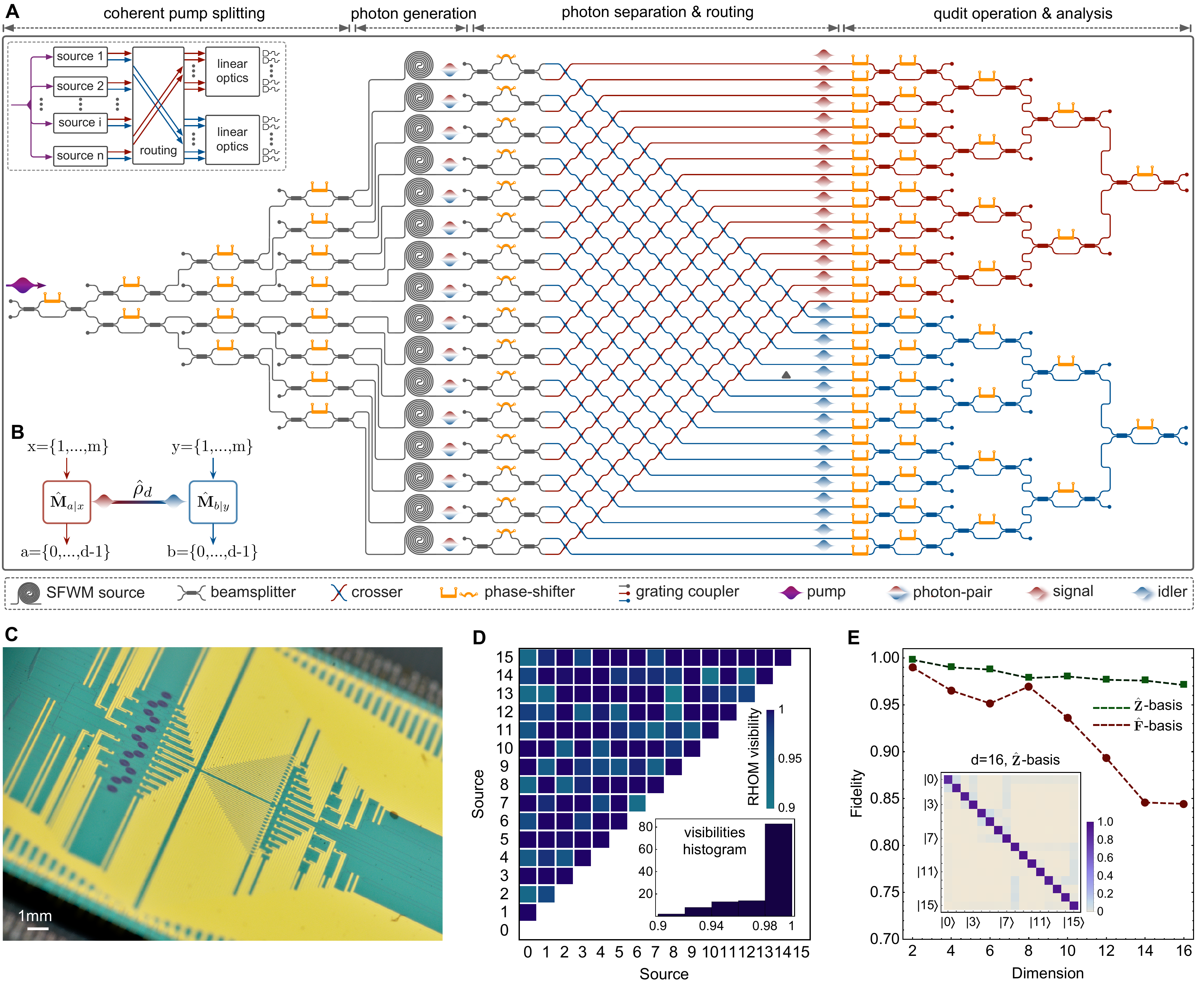} 
\caption{\textbf{Diagram and characterization of the multidimensional silicon quantum photonic circuit.}  
(\textbf{A}) Circuit diagram. The device monolithically integrates 16 SFWM photon-pair sources, 93 thermo-optical phase-shifters, 122 multimode interferometers (MMI) beamsplitter, 256 waveguide-crossers and 64 optical grating couplers. 
A photon pair is generated by SFWM in superposition across 16 optical modes, producing a tunable multidimensional bipartite entangled state. The two photons, signal and idler, are separated by an array of asymmetric MZI filters and routed by a network of crossers, allowing the local manipulation of the state by linear-optical circuits. Using triangular networks of MZIs, we perform arbitrary local projective measurements. 
The inset represents a general schematic for universal generation and manipulation of bipartite multidimensional entangled  states. 
(\textbf{B}) Framework for correlation measurements on a shared $d$-dimensional state $\hat{\rho}_d$. $\hat{M}_{a|x}$ and $\hat{M}_{b|y}$ represent the operators associated to local measurements  $x$ on Alice and $y$ on Bob, with outcomes $a$ and $b$ respectively. 
(\textbf{C}) Photograph of the device. Silicon waveguides and 16 SFWM sources 
can be observed as black lines. Gold wires allow the electronic access of each phase-shifter. 
(\textbf{D}) Visibilities for the two-photon RHOM experiments to test sources' indistinguishability.  
The inset shows the histogram of all 120 measured visibilities, with a mean value of $0.984 \pm 0.025$. 
(\textbf{E}) Statistical fidelity for $d$-dimensional projectors, in both the computational $\hat{Z}$-basis and the Fourier $\hat{F}$-basis. 
The inset shows the measured distribution for the $16$-dimensional projector in the $\hat{Z}$-basis. 
}
\label{fig:Schematic}
\end{figure*}    
 
\subsubsection*{Large-scale integrated quantum photonic circuit}

\begin{figure*}[ht!]  
\centering
\includegraphics[width=1\textwidth]{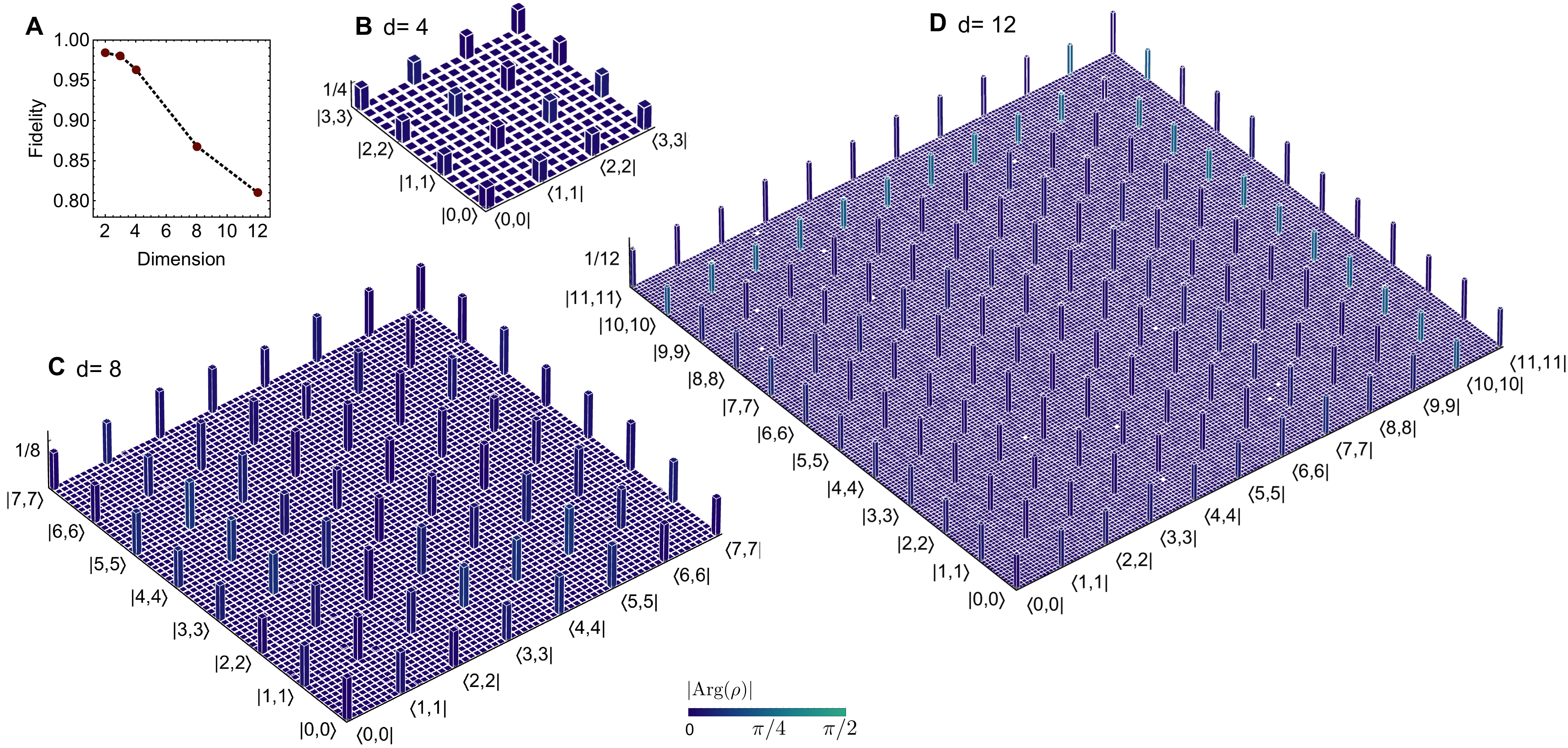} 
\caption{\textbf{Experimental quantum state tomographies.}  
(\textbf{A}) Measured quantum fidelities $\bra{\psi^+_d}\hat{\rho}_d\ket{\psi^+_d}$, where  $\hat{\rho}_d$ represents the reconstructed states and $\ket{\psi^+_d}$ refers to the ideal $d$-dimensional maximally entangled state. 
Reconstructed density matrices for the entangled states in dimension (\textbf{B}) 4, (\textbf{C}) 8, and (\textbf{D}) 12 using compressed sensing techniques.  
Column heights represent the absolute values $|\rho|$ while colors represent the phases $|\text{Arg}(\rho)|$. 
The phase information for matrix elements with module $|\rho_{ij}|<0.01$ is approximately randomly distributed and not displayed for more clarity. 
} 
\label{fig:Tomos}
\end{figure*}

Entangled path-encoded qubits can be generated by coherently pumping two spontaneous parametric down conversion~\cite{Schaeff:Mode, Ciampini:2016bs} or spontaneous four-wave mixing (SFWM) photon-pair sources~\cite{Silverstone2013,Wang2016}.
The approach can be generalized to qudits via the generation of photons entangled over $d$ spatial modes by coherently pumping $d$ sources~\cite{Schaeff:Mode,Krenn2017}. However, scaling this approach to high dimensions has represented a significant challenge, due to the need of a stable and scalable technology able to coherently embed large arrays of identical photon sources and to precisely control qudit states in large optical interferometers. 

Silicon quantum photonics, offering intrinsic stability~\cite{Silverstone2013,Bonneau2016}, high precision~\cite{Wang:QHL,BPEA} and dense integration~\cite{Harris,Sun:2013kl},
can provide a natural solution. 
In this work we devise a large-scale silicon quantum photonic circuit to implement the scheme, as shown in Fig.~\ref{fig:Schematic}A.  
A total of 16 SFWM sources are coherently pumped, generating a photon-pair in a superposition across the array. 
As both the photons must originate from the same source, the bipartite state created is 
$\sum_{k=0}^{d-1} c_{k}\ket{1}_{i,k}\ket{1}_{s,k}$
where $\ket{1}_{i,k}$ ($\ket{1}_{s,k}$) indicates the Fock state of the idler (signal) photon being in its $k$-th spatial mode and $c_k$ represents the complex amplitude in each mode (with $\sum |c_k|^2 =1$). 
The mapping between the Fock state of each photon and the logical state is the following: we say that the qudit state is $\ket{k}$ ($k=0,\ldots,d-1$) if the associated photon is in its $k$-th optical mode.  
This yields an arbitrary multidimensional entangled state:
\begin{equation}
|\psi\rangle_d=\sum_{k=0}^{d-1} c_k 
\ket{k}_i\ket{k}_s,
\label{eq:state}
\end{equation}
where the coefficients $c_k$ can be arbitrarily chosen by controlling the pump distribution over the $d$ sources and the relative phase on each mode.
This is achieved using a network of Mach-Zehnder interferometers (MZIs) at the input and phase-shifters on each mode, as shown in  Fig.~\ref{fig:Schematic}A. 
In particular, maximally entangled states $|\psi^+_d\rangle=\sum_{k=0}^{d-1} \ket{k}_i\ket{k}_s / \sqrt{d}$ can be obtained with a uniform excitation of the sources. 
The two non-degenerate photons are deterministically separated using asymmetric MZI filters and routed by a network of waveguide crossings, grouping the signal photon into the top modes and the idler photon into the bottom ones (see Fig.~\ref{fig:Schematic}A). 
We can then locally manipulate and measure the state of each qudit. Linear-optical circuits enable the implementation of any local unitary transformation $\hat{U}_d$ in dimension $d$
~\cite{Reck,Oxford:U,Carolan:15}. 
Here we use a triangular network of MZIs and phase-shifters, shown in Fig.~\ref{fig:Schematic}A, which allows us to 
perform arbitrary local projective measurements. For more details on the device and the experimental setup see Supplementary Materials 1. 

The 16 photon-pair sources are designed to be identical. 
Two-photon reversed Hong-Ou-Mandel (RHOM\xspace) interference is used to verify their performance, where the fringe visibility gives an 
estimate of the sources' indistinguishability~\cite{Silverstone2013}. 
RHOM interference is tested between all the possible pairs of the 16 sources, performing $\binom{16}{2}=120$ quantum interference experiments and evaluating the corresponding visibilities. 
The pair of sources used for each interference experiment is selected each time by reconfiguring the interferometric network. Approximately a 2kHz photon-pair detection rate is observed in typical measurement conditions. 
In Fig. \ref{fig:Schematic}D the measured visibilities are reported. 
In all cases, we obtained a visibility $\text{>}0.90$, and more than $80\%$ cases presented $\text{>}0.98$ visibility. These results show a state-of-the-art degree of source indistinguishability in all 120 RHOM experiments, leading to the generation of high quality entangled qudit states.  

Each of the MZIs and phase-shifters can be rapidly reconfigured (kHz rate) with high precision~\cite{Bonneau2016,Harris}. 
The quality of the qudit projectors is characterized by the classical statistical fidelity, which quantifies the output distribution obtained preparing and measuring a qudit on a fixed basis. 
As reported in Fig.~\ref{fig:Schematic}E, we measured the fidelity of projectors in dimension $d=2$ to $16$ in both the computational basis $\hat{Z}=\ket{k}\bra{k}$, and in the Fourier-transform basis $\hat{F}=\ket{\ell}\bra{\ell}$, where $\ket{\ell}=\sum_{k=0}^{d-1} e^{2\pi i k \ell/d } \ket{k}/\sqrt{d}$ and $k,\ell=0,\ldots,d-1$. 
We observe for $d=8$ fidelities of $98\%$ in the $\hat{Z}$-basis and $97\%$ in the $\hat{F}$-basis, while for $d=16$ fidelities of $97\%$ in the $\hat{Z}$-basis and $85\%$ in the $\hat{F}$-basis. More details are provided in Supplementary Materials 1. 
The residual imperfections are mainly due to thermal cross-talk between phase-shifters (higher in the $\hat{F}$-basis), which can be mitigated using optimized designs for the heaters~\cite{Harris} or ad-hoc characterization techniques~\cite{Carolan:15,BPEA}.

Due to a fabrication imperfection in the routing circuit one of the modes (triangle label in Fig.~\ref{fig:Schematic}A) for the idler photon presents an additional 10 dB loss. For simplicity we exclude this lossy mode in the rest of our experiments, and study multidimensional entanglement for dimension up to $15$.

Figure~\ref{fig:Schematic}B represents the experiment in the standard framework for bipartite correlation. The correlations between two parties Alice (A) and Bob (B), here identified by the signal and idler photon respectively, are quantified by joint probabilities $p(ab|xy)=\text{Tr}[\hat{\rho}_d (\hat{M}_{a|x} \otimes \hat{M}_{b|y})]$, where $\hat{\rho}_d$ is the shared $d$-dimensional state, $x,y \in \{1,\ldots,m\}$ represent the $m$ measurement settings chosen by Alice and Bob, and $a,b \in \{ 0,\ldots,d-1\}$ label the possible outcomes with associated measurement operators  $\hat{M}_{a|x}$ and $\hat{M}_{b|y}$.

\subsubsection*{Quantum state tomographies} 

Quantum state tomography (QST) allows us to estimate the full state of a quantum system, providing an important diagnostic tool.  
In general, performing a complete tomography is a very expensive task both in terms of the number of measurements and the computational time to reconstruct the density matrix from the data. For these reasons complete QST on entangled qudits states has been achieved only up to 8-dimensional systems~\cite{Agnew2011}. 
In order to perform the tomographic reconstructions of larger entangled states, we use quantum compressed sensing techniques. Inspired by advanced classical methods for data analysis, these techniques significantly reduce the experimental cost for state reconstruction~\cite{Gross:2010cv}, 
are general for density matrices of arbitrary dimension~\cite{Gross2011,Bolduc:2016fv}, and have been experimentally demonstrated to characterize complex quantum systems~\cite{CompressTomo,Bolduc:2016fv}. We experimentally implement compressed sensing QST to reconstruct  bipartite entangled states with local dimension up to $d=12$. Fidelities with ideal states $|\psi^+_d\rangle$ are reported in Fig.~\ref{fig:Tomos}A. For dimensions $d=4,\ 8$ and $12$ we plot in Fig.~\ref{fig:Tomos} the reconstructed density matrices, with fidelities of $96\%$, $87\%$ and $81\%$, respectively. 
These results show a significant improvement of the quality for multidimensional entanglement~\cite{Agnew2011,Roberto:frequency}. 
More details are reported in Supplementary Materials 2.

\subsubsection*{Certification of system dimensionality}

\begin{figure}[t!]
\centering 
\includegraphics[width=0.46\textwidth]{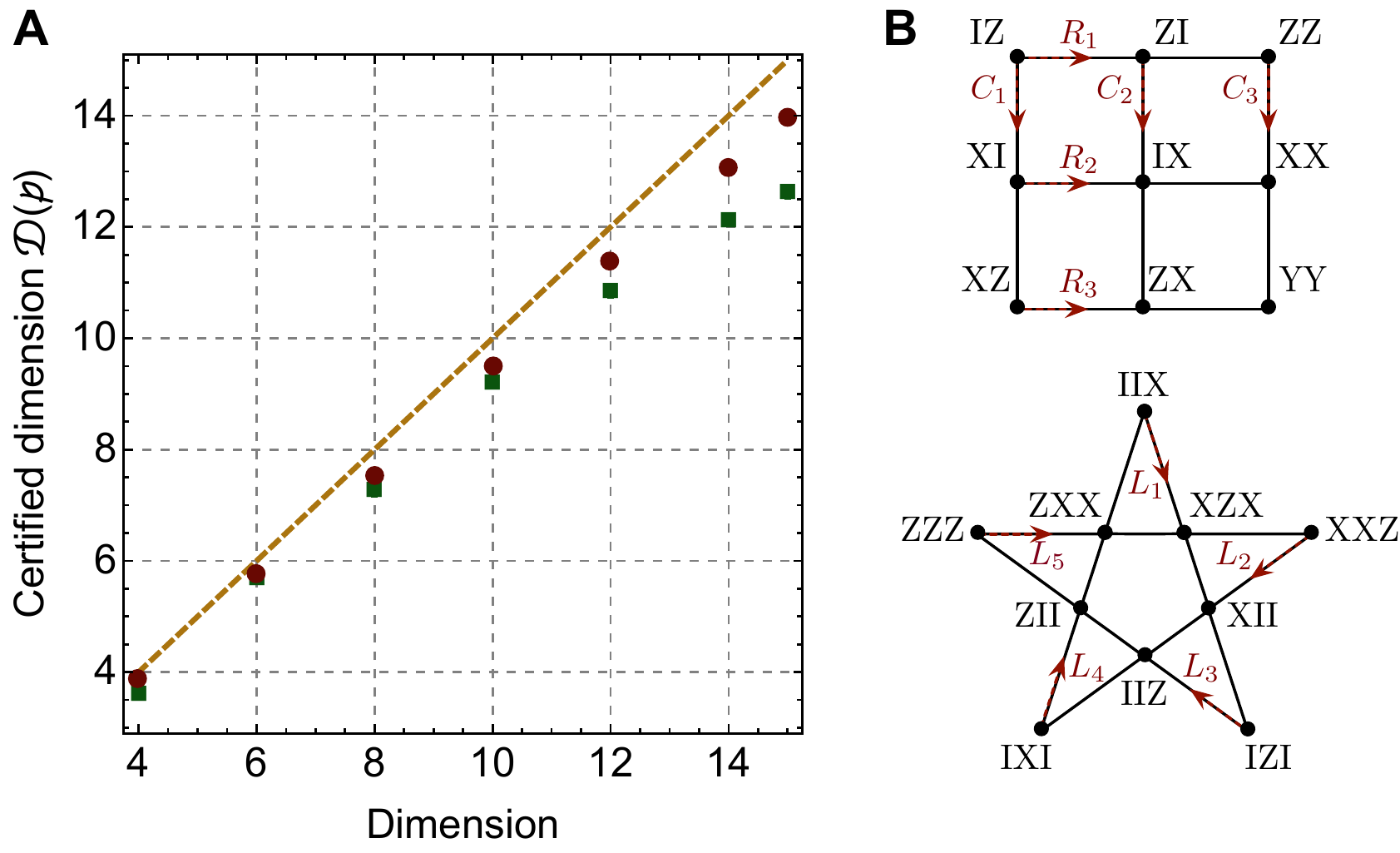}
\caption{\textbf{Verification of system dimensionality. 
} 
(\textbf{A}) Experimental results. Data points refer to the measured lower bounds on the local dimension of the generated entangled states; green (red) points represent data for the measurement scenario I (II). The yellow line refers to ideal values. 
Errors are smaller than the markers and neglected in the plot for clarity. 
(\textbf{B}) Correlation measurements 
associated to optimal strategies for Magic Square and Pentagram games. 
$X$, $Y$ and $Z$ are Pauli operators and $I$ is the identity. Red lines $C_i$, $R_i$ and $L_i$ are associated to different measurement settings.
Single Magic Square game, Magic Pentagram game, and two copies of the Magic Square game are used to certify the dimension for states with $d\leq4$, $4< d\leq 8$, and $8<d\leq 15$, respectively.  }
\label{fig:DW}
\end{figure}

The dimension of a quantum system quantifies its ability to store information and represents a key resource for quantum applications.  
Device-independent (DI) dimension witnesses enable us to lower bound the dimension of a quantum system solely from the observed statistics, \textit{i.e.}, correlation probabilities $p(ab|xy)$, making no prior assumptions on the experimental apparatus (see e.g.~\cite{Brunner08Testing,Navascues15Testing,Sikora:DIDW}). 
Here, we adopt the approach of Ref.~\cite{Sikora:DIDW} to verify the local dimension of entangled states in a DI way in the context where shared randomness is not a free resource. 
The lower bound takes the form of $d\geq \lceil \mathcal{D}(p) \rceil$, where $\mathcal{D}(p)$ is a nonlinear function of the correlations, 
and $\lceil \epsilon \rceil$ indicates the least integer $\geq \epsilon$. 
We adopt two different measurement scenarios, with experimental results shown in Fig.~\ref{fig:DW}A. In scenario I, we calculate the bound from the measured (partial) correlations for the Magic Square and Pentagram games~\cite{Mermin1990}, shown in Fig.~\ref{fig:DW}B. 
For example, to certify $8\times 8$ entangled states, locally equivalent to a 3-qubit system, we perform a $\hat{Z}$-basis measurement on Alice's system, while on Bob's system we use the $\hat{Z}$-basis and the one which simultaneously diagonalizes the commuting operators $ZZZ,\ ZXX,\ XZX$, and $XXZ$ (see lines L3 and L5 in Fig.~\ref{fig:DW}B respectively).   
In the absence of noise we would achieve $\mathcal{D} =8$. Using the measured correlations we obtain $\mathcal{D}(p_8^\text{I}) \ge 7.22 \pm 0.05$ which yields the optimal lower bound $\lceil \mathcal{D}(p_8^\text{I}) \rceil=8$. 
In scenario II, we compute $\mathcal{D}(p_d^\text{II})$ for correlations $p_d^\text{II}$ obtained by performing $\hat{Z}$-basis measurements on both sides of the maximally entangled state of local dimension $d$. We expect less experimental noise in this scenario (see Fig.~
\ref{fig:Schematic}D). 
As shown in Fig.~\ref{fig:DW}A, the experimentally observed correlations $p_d^{\text{II}}$ yield $\lceil \mathcal{D}(p_d^\text{II}) \rceil = d$ for all $d\le 14$, certifying the correct dimensions. Further details can be found in Supplementary Materials 3.

\subsubsection*{Multidimensional Bell non-locality and state self-testing}

 \begin{figure*}[ht!]
\centering 
\includegraphics[width=0.66\textwidth]{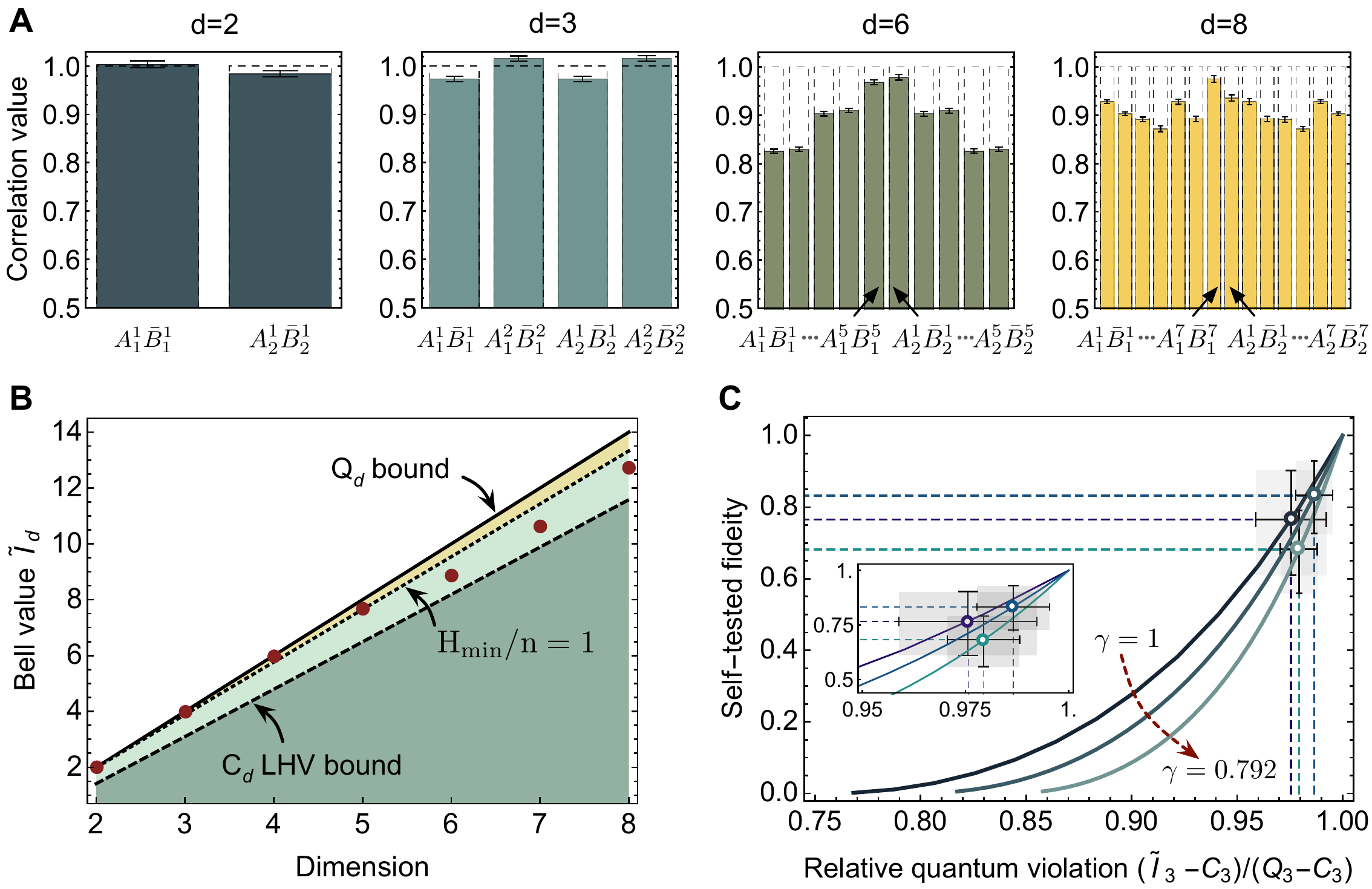}
\caption{\textbf{
Bell violation and self-testing on multidimensional entangled states.}  
(\textbf{A}) Measured values of the 2($d-1$) correlators $\text{Re}\small[\langle {A^l_i} {\bar{B}^l_i} \rangle\small]$.  
Dashed boxes refers to theoretical values. 
(\textbf{B}) Violation of the generalized SATWAP Bell-type inequalities for $d$-dimensional states. Red points are experimentally measured  $\tilde{I}_d$ values. Non-locality is certified by violating $\tilde{I}_d\leq C_d$, where $C_d$ is the classical LHV bound (dashed line). The Tsirelson bound $Q_d$ (solid line) represents the maximal violation for quantum systems.  
The dotted line represents the threshold above which more than $1$ global random bit can be extracted per output symbol from Bell correlations. 
(\textbf{C}) DI self-testing of entangled qutrit states $(\ket{00}+\gamma \ket{11}+\ket{22})/\sqrt{2+\gamma^2}$ for $\gamma=1,~0.9,~ 0.792$. 
Self-tested minimal fidelities to ideal states are plotted as a function of the relative violation for more clarity.
The significant uncertainty on the fidelity value is due to the general limited robustness of self-testing protocols. All errors are estimated from photon Poissonian statistics, and those in (\textbf{B}) are smaller than markers.    
}
\label{fig:Bell}
\end{figure*}

Bell inequalities enable to experimentally study quantum non-locality, which indicates the presence of correlations incompatible with local-hidden variables (LHV) theories. 
Non-locality can be demonstrated by the violation of Bell inequalities of the form $S_d \leq C_d$, where the parameter $S_d$ is a linear function of the joint probabilities, and $C_d$ is the classical bound for LHV models.  
We study two types of generalized Bell-type inequalities for $d$-dimensional bipartite systems:  the SATWAP inequalities (Salavrakos--Augusiak--Tura--Wittek--Acin--Pironio), recently introduced in Ref.~\cite{SATWAP}, and the standard CGLMP inequalities (Collins--Gisin--Linden--Massar--Popescu)~\cite{CGLMP}. 
In contrast to CGLMP inequalities, SATWAP inequalities are explicitly tailored to obtain a maximal violation for maximally entangled qudit states. 
Here we test the 2-input version of the SATWAP inequalities by measuring the joint probabilities to obtain the quantity
\begin{equation}
\tilde{I}_{d}=\sum_{\substack{i=1} }^{2}\sum_{\substack{l=1}}^{d-1} \langle {A^l_i} {\bar{B}^l_i} \rangle, 
\label{eq:SATWAPequ}
\end{equation}
where the $2(d-1)$ values $\langle {A^l_i} {\bar{B}^l_i} \rangle$ represent generalized Bell correlators, whose explicit form is given in Supplementary Materials 4. The Bell inequality here is given by $\tilde{I}_{d}\leq C_d$, where the bound for classical LHV models is $C_d=\[3\cot(\pi/4d)-\cot(3\pi/4d)\]/2-2$. 
The maximum value of $\tilde{I}$ obtainable with quantum states (Tsirelson bound) is known analytically for arbitrary dimensions and is given by $\tilde{I}_{d}\leq Q_d=2d-2$. This maximal violation is achieved with maximally entangled states~\cite{SATWAP}.

In Fig.~\ref{fig:Bell}A we show the experimental values of the generalized correlators $\text{Re}\small[\langle {A^l_i} {\bar{B}^l_i} \rangle\small]$. 
The correlation measurements are performed in the Fourier bases provided in Supplementary Materials 4. 
Figure~\ref{fig:Bell}B shows the obtained values of $\tilde{I}_d$ for dimensions $2$ to $8$, together with the analytical quantum and classical bounds. 
In all cases the classical bound is violated. In particular in dimensions  2--4 a strong violation is observed, closely approaching the Tsirelson bound $Q_d$. 

We report in Table~\ref{t:BELL} the experimental values for the CGLMP inequalities. Also for CGLMP, strong violations of LHV models are observed. As an example, for $d=4$ we observe $S_4=2.867\pm 0.014$, 
which violates the classical bound (i.e. $C_d=2$ for CGLMP inequalities) by $61.9\ \sigma$, and is higher than the maximal value achievable by 2-dimensional quantum systems ($S_2=2\sqrt[]{2}$) by $2.8\ \sigma$, indicating stronger quantumness for higher dimensions. 

The near-optimal Bell violations enable the self-testing of multidimensional entangled states. Self-testing represents the characterization of quantum devices without assumptions on their inner functioning. As is desirable for practical quantum applications, it allows classical users to certify quantum devices without any prior knowledge. This fully DI characterization can be obtained based solely on observed non-local correlations~\cite{MayersYao,Coladangelo2017}. In more details, if the maximal violation of a Bell inequality can only be achieved by a unique quantum state and set of measurements (up to local isometries), a near-optimal violation enables to characterize the experimental apparatus. 
In Ref.~\cite{SATWAP} it was shown that the SATWAP inequality can be used to self-test the maximally entangled state of two qutrits $|\psi_3^+\rangle$; %
in particular, employing a numerical approach from Ref.~\cite{Yang2014}, a lower bound on the state fidelity can be obtained from the measured value of $\tilde{I}_3$.
In  Supplementary Materials 4 we generalize it also for arbitrary qutrit states of the form $\ket{00}+\gamma \ket{11}+\ket{22}$ (up to normalization). In Fig.~\ref{fig:Bell}C we report the experimental self-tested fidelities for different values of $\gamma=1,\  0.9,\text{and } (\sqrt{11}-\sqrt{3})/2\approx 0.792$. This is possible by exploiting the capability of the device to generate multidimensional states with tunable entanglement. In particular, $\gamma=1$ indicates $|\psi_3^+\rangle$ and $\gamma=(\sqrt{11}-\sqrt{3})/2$ represents the state that maximally violates the CGLMP inequality~\cite{Yang2014}. We experimentally achieve an average self-tested fidelity of 77\%. We remark that the certification of high fidelities in a self-testing context is only achievable in the presence of near-ideal experimental correlations. The measured self-tested fidelities are comparable with the reported values obtained from full tomographies in other experimental approaches~\cite{Agnew2011,Roberto:frequency}. Although our device provides high violations also for dimensions higher than 3, it remains an open problem whether the approach based upon SATWAP inequalities can be generalized to self-test states in arbitrary dimension~\cite{SATWAP}.

\begin{table}
\centering
\begin{tabular}{c | r@{$\pm$}l r@{$\pm$}l } 
 \hline \hline \noalign{\smallskip} 
Dim
& \multicolumn{2}{c}{CGLMP $S_d$}
& \multicolumn{2}{c}{SATWAP $\tilde{I}_d$}\\
\hline \noalign{\smallskip}
$2$ & ($2$)~$2.810$&$ 0.014$~\{$2.828$\} & ($1.414$)~$1.987 $&$ 0.010$~\{$2$\}\\
 $3$ & ($2$)~$2.845$&$ 0.012$~\{$2.873$\} & ($3.098)$~$3.978 $&$ 0.015$~\{$4$\}\\
$4$ & ($2$)~$2.867$&$0.014$~\{$2.896$\} & ($4.793$)~$5.978 $&$ 0.032$~\{$6$\}\\
$5$ & ($2$)~$2.763$&$0.014$~\{$2.910$\} & ($6.489$)~$7.652 $&$ 0.031$~\{$8$\}\\
$6$ & ($2$)~$2.629$&$0.010$~\{$2.920$\} & ($8.187$)~$8.883 $&$ 0.029$~\{$10$\}\\
 $7$ & ($2$)~$2.532$&$0.013$~\{$2.927$\} & ($9.884$)~$10.645 $&$ 0.029$~\{$12$\}\\
$8$ & ($2$)~$2.650$&$0.012$~\{$2.932$\} & ($11.581$)~$12.740 $&$ 0.044$~\{$14$\}\\ \hline
\end{tabular}
\caption{\label{t:BELL}\textbf{Experimental values for multidimensional Bell correlations}. Measured CGLMP and SATWAP values are given with experimental errors. Values in (*) refer to the LHV classical bound; those in \{*\} refer to theoretical bounds for $d$-dimensional maximally entangled states. Errors are given by photon Possionian noise. 
}
\end{table}

\begin{figure}[ht!] 
\centering  
\includegraphics[width=0.31\textwidth]{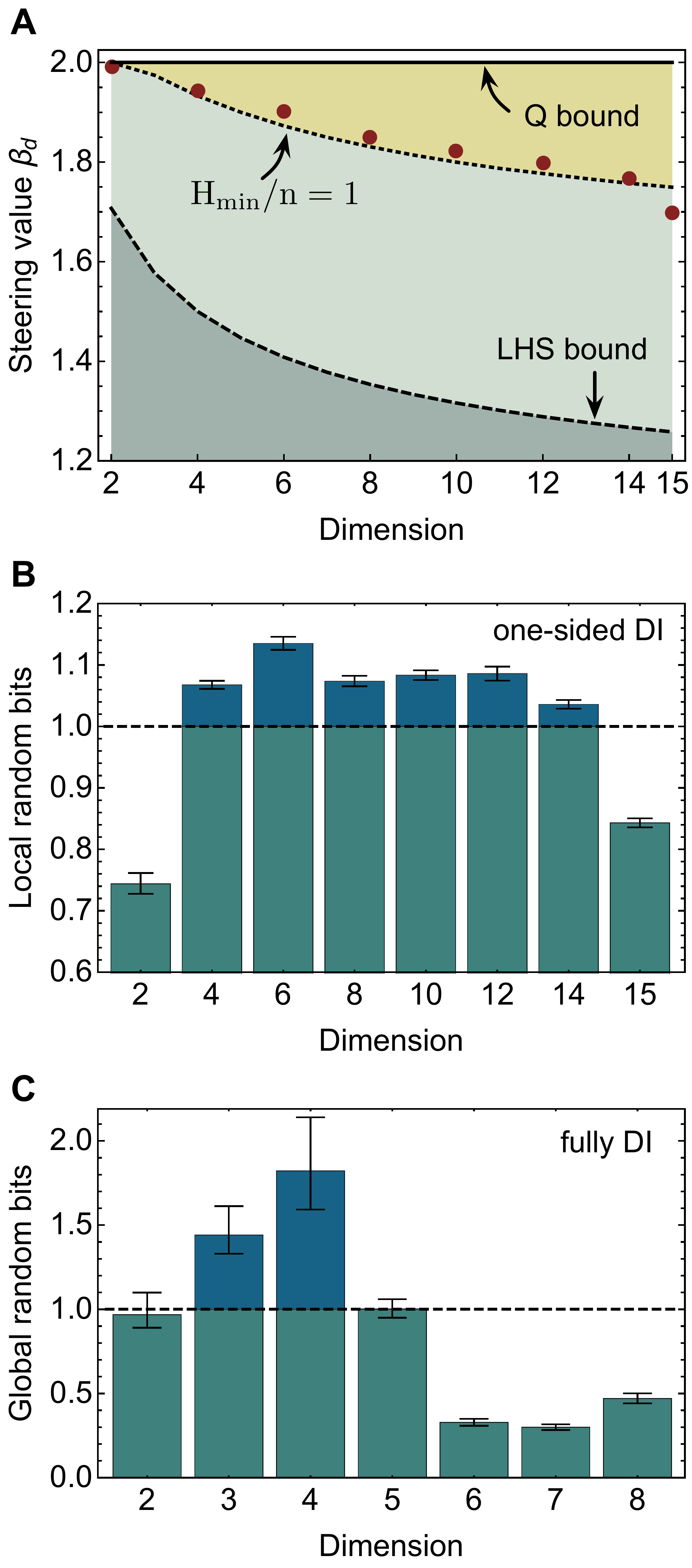}
\caption{\textbf{Certification of multidimensional randomness expansion.} 
(\textbf{A}) Multidimensional EPR steering is certified by violating the inequality $\beta_d\leq \beta_{lhs}$, where $\beta_{lhs}$ is the classical bound for LHS models (dashed line). 
Red points are experimentally measured steering values $\beta_d$. 
The dotted line denotes the threshold above which more than $1$ local random bit can be extracted per output symbol from steering correlations. 
(\textbf{B}) Local randomness per symbol certified in a one-sided DI scenario by $d$-dimensional steering correlations. (\textbf{C}) Global randomness per symbol certified in a fully DI scenario by $d$-dimensional Bell correlations. 
Above the dashed line in (\textbf{B}) and (\textbf{C}), more than 1 private random bits are generated. 
Error bars are given by Poissonian statistics, and those in (\textbf{A}) are smaller than markers. 
}
\label{fig:Steering}
\end{figure}

\subsubsection*{Multidimensional randomness expansion} 
Randomness is a key resource in many practical applications. 
Generating randomness is however a notoriously difficult problem. 
Quantum theory, being fundamentally nondeterministic, provides a natural solution. The unpredictability of measurement outcomes forms the basis of quantum random number generators \cite{RevModPhys.RNG,,Acin:2016ke}. Remarkably, quantum theory can go one step further: the presence of non-local correlations can certify unconditional randomness of measurement outcomes even without full knowledge of the experimental apparatus used~\cite{Pironio:2010}. There are a number of different scenarios where this is possible, which include the strongest fully-DI scenario, based on Bell non-locality, and the one-sided DI (1SDI) scenario, based on EPR steering. In the former, neither the source of particles nor any of the measuring devices 
are trusted, while in the latter one of the measuring devices 
is assumed to be characterized ~\cite{SteeringRAM,Passaro2015}. 

We use the above SATWAP Bell inequalities to study DI randomness expansion~\cite{RevModPhys.RNG}, while in the 1SDI case we use the EPR steering inequality
(see Ref.~\cite{Cavalcanti2017} and Supplementary Materials 5)
\begin{equation}
\beta_d=\sum_{\substack{a=b \\ x=y}}p(a|x)\text{Tr}[\hat{M}_{b|y}\hat{\rho}_{a|x}].
\label{eq:steering}
\end{equation}
Here $p(a|x)$ are the probabilities of Alice's uncharacterized measurements; $\hat{M}_{k|0}=\ket{k}\bra{k}$, $\hat{M}_{\ell|1}=\ket{-\ell}\bra{-\ell}$ are the characterized measurements of Bob, with $\ket{k}$ corresponding to the $\hat{Z}$-basis and $\ket{\ell}$ to the $\hat{F}$-basis, defined above; $\hat{\rho}_{a|x}$ indicates the reduced state for Bob when the measurement $x$ is performed on Alice and  outcome $a$ is obtained. 
In any classical local hidden state (LHS) model $\beta_d$ is bounded by $\beta_d\leq \beta_{lhs}=1 + 1/\sqrt{d}$, while quantum states can violate this inequality and maximally achieve $\beta_d = 2$. 
Figure~\ref{fig:Steering}A reports the measured values of $\beta_d$  up to dimension $15$, all violating the LHS classical bound. Note that, as steering is only possible if the underlying shared bipartite state is entangled~\cite{PRL_Doherty}, this also certifies, in a 1SDI manner, that all the generated states are entangled up to $d=15$.    

The idea behind DI randomness certification is that a party who wants to generate private randomness performs $n$ runs of a Bell or steering test using a source of (public or private) randomness to choose the measurement settings. 
Based on the observed violation of the corresponding inequality, the predictability of the outcomes string by a third party (Eve) can be upper bounded, and private randomness is thus generated. 
Randomness expansion is, by definition, achieved if more private random bits are generated than those consumed to choose the measurement settings.  

We study here the case where randomness is generated from both measuring devices in the DI setting (global randomness), and from the uncharacterised device in the 1SDI setting (local randomness). In the former case randomness is quantified in terms of the \textit{min-entropy} $H_{min}(AB|E) =-\log_2 P_g(AB)$, where $P_g(AB)$ is the predictability for Eve of the pair of outcome strings for Alice and Bob, while in the latter case $H_{min}(A|E) = - \log_2 P_g(A)$, where $P_g(A)$ is the predictability for Eve of the outcome string of Alice. DI and 1SDI bounds can be placed on these quantities \cite{Pironio:2010,CavSkr17InPrep}. 
Details are provided in Supplementary Materials 6.

A particularly demanding task is the efficient generation of randomness -- to generate more than 1 bit of randomness per output symbol, i.e. to achieve $H_{min}(AB|E) > n$ or $H_{min}(A|E) > n$. In this regime randomness expansion is naturally achieved as more than one private random bit is obtained per round. 
For qubits, this is only possible using non-projective measurements (with more than 2 outcomes) \cite{Acin16} or with sequences of measurements~\cite{Curchod17}. In contrast, multidimensional entangled states provide a natural route, based upon projective measurements, where up to $n\log_2 d$ bit of randomness can be expected in the ideal case. Numerical investigation shows that this upper limit is certified by a maximal violation of the SATWAP Bell inequality Eq.~\eqref{eq:SATWAPequ}, while for the steering inequality Eq.~\eqref{eq:steering} this can be shown analytically \cite{CavSkr17InPrep}.

In Figures~\ref{fig:Bell}B and~\ref{fig:Steering}A the minimum values of $\bar{I}_d$ and $\beta_d$ above which more than one bit of (global or local) randomness per symbol is certified in a DI or one-sided DI setting, are reported for different dimensions (yellow regions). 
The global randomness associated with the Bell violations shown in Fig.~\ref{fig:Bell}A, are reported in Fig.~\ref{fig:Steering}C. Efficiency $H_{min}/n>1$ is achieved for $d = 3$ and $4$. The largest amount of randomness per symbol is obtained for $d=4$, where $H_{min}/n=1.82 \pm 0.35$ random bits. The experimentally measured values of $\beta_d$ are shown in Fig.~\ref{fig:Steering}A, and the associated local randomness is reported in Fig.~\ref{fig:Steering}B. Here efficiency $H_{min}/n>1$ is preserved for the range $4\leq d \leq 14$, indicating, as expected, stronger robustness in the 1SDI case.

\subsubsection*{Conclusion}
We have shown how silicon-photonics quantum technologies
have reached the maturity level which enables fully on-chip generation, manipulation and analysis of multidimensional quantum systems. 
The achieved complexity of our integrated device
represents a significant step forward for large-scale quantum photonic technologies, opening the door to a wide range of practical applications. For example, high-rate device-independent randomness generators can be realized harnessing the abilities of efficient randomness expansion shown here and high-speed on-chip state manipulation~\cite{Reed:im}. 
Together with recently developed techniques for inter-chip state distribution~\cite{LaingRFIQKD,Wang2016,Ding:QKD}, our approach can lead to the future development of high-dimensional chip-based quantum networks (see preliminary results in Supplementary Materials $7$). 
Moreover, the scalability of silicon photonics can further increase system dimensionality, and allow the coherent control of 
multiple photons entangled over a large number of modes. 
Our results pave the way for the development of advanced multidimensional quantum technologies.

\printbibliography

\section*{Acknowledgments} 
We acknowledge A.C. Dada, P.J. Shadbolt, J. Carolan, C. Sparrow, W. McCutcheon, A.A. Gentile, D.A.B. Miller and Q. He for useful discussions. We thank W.A. Murray, M.Loutit  and R.Collins for experimental assistance.  
\textbf{Funding:} we acknowledge the support from the Engineering and Physical Sciences Research Council (EPSRC), European Research Council (ERC), and European Commission (EC) funded grants PICQUE, BBOI, QuChip, QITBOX, and the Center of Excellence, Denmark SPOC (ref DNRF123), Bristol NSQI. 
J.W. acknowledges the Chinese National Young 1000 Talents Plan. 
P.K. is supported by the Royal Society through a University Research Fellowship (UHNL). 
L. M. is supported by the Villum Fonden via the QMATH Centre of Excellence (Grant No. 10059) and also acknowledges the support from the EPSRC (Grant EP/L021005/1). 
We also acknowledge support from the Spanish MINECO (QIBEQI FIS2016-80773-P and Severo Ochoa SEV-2015-0522), the Fundacio Cellex, the Generalitat de Catalunya (SGR875 and CERCA Program), and the AXA Chair in Quantum Information Science. 
This project has received funding from the European Union's Horizon 2020 research and innovation programme under the Marie Sk{\l}odowska-Curie grant agreements no. 705109, 748549, and 609405. 
Q.G. acknowledges the National Key R and D Program of China (no.2013CB328704). 
J.L.O. acknowledges a Royal Society Wolfson Merit Award and a Royal Academy of Engineering Chair in Emerging Technologies.  
Fellowship support from EPSRC is acknowledged by A.L. (EP/N003470/1). M.G.T. acknowledges support from the ERC starter grant ERC-2014-STG 640079 and and an EPSRC Early Career Fellowship EP/K033085/1. 
\textbf{Authors contributions:} 
J.W., S.P. and Y.D. contributed equally to this work. J.W. designed the experiment. Y.D. designed and fabricated the device. J.W., S.P., Y.D., R.S. and J.W.S. built the setup and carried out the experiment. S.P, P.S., A.S., J.T., R.A., L.M., D.B. and D.Bonneau performed the theoretical analysis. Q.G., A.A., K.R., L.K.O., J.L.O., A.L and M.G.T. managed the project. All authors discussed the results and contributed to the manuscript.

\newpage 
\clearpage
\pagenumbering{arabic}
\setcounter{page}{1}

\onecolumn
\section*{\centering\fontsize{15}{15}\selectfont 
Supplementary Materials: \\
Multidimensional Quantum Entanglement with Large-scale Integrated Optics}

\begin{center}
\begin{minipage}{1\textwidth}
\begin{center}
\author{\large{Jianwei Wang$^{1,2\dagger}$, Stefano Paesani$^{1\dagger}$, Yunhong Ding$^{3,4\dagger}$, Raffaele Santagati$^{1}$, Paul Skrzypczyk$^{5}$, Alexia Salavrakos$^{6}$, Jordi Tura$^{7}$, Remigiusz Augusiak$^{8}$, Laura Man\v{c}inska$^{9}$, Davide Bacco$^{3.4}$, Damien Bonneau$^{1}$, Joshua W. Silverstone$^{1}$, Qihuang Gong$^{2}$, Antonio Ac\'in$^{6,10}$, Karsten Rottwitt$^{3,4}$, Leif K. Oxenl{\o}we$^{3,4}$, Jeremy L. O\textquoteright Brien$^{1}$, Anthony Laing$^{1}$, Mark G. Thompson$^{1}$}} 
\end{center}
\end{minipage}
\end{center}

\section{Device and experimental setup details}
\subsection{Device fabrication and components characterization}
The silicon quantum photonic integrated circuit is designed and fabricated on commercial silicon-on-insulator (SOI) platform with top silicon thickness of 260 nm and buried oxide (BOX) layer of 1 {\micro\metre}. 
First, e-beam lithography (Ebeam writer JBX-9500FSZ) is used to make the resist mask pattern on the SOI wafer. Afterwards, inductively coupled plasma (STS ICP Advanced Silicon Etcher) etching is applied to transfer the pattern from the resist mask layer to silicon. Then a 1500 nm thick layer of $\text{SiO}_2$ is deposited on top of the chip by plasma-enhanced chemical vapor deposition. The chip surface is polished afterwards, and the top $\text{SiO}_2$ is thinned down to 1 {\micro\metre} accordingly. The 1 {\micro\metre} $\text{SiO}_2$ is used as an isolation layer between the silicon waveguides and the Ti heaters to avoid potential optical losses. After that, a second e-beam lithography is used to define the patterns for Ti heaters, which are formed later by 100nm Ti deposition followed by metal liftoff process. UV lithography (Aligner MA6) is utilized to define the patterns of contact wires and pads, which are fabricated afterwards by thick Au/Ti deposition and liftoff process. Finally, the chip is cleaved and wire-bonded to a PCB board. 

The silicon nanophotonic waveguides are designed and fabricated with a size of $450~\text{nm} \times 260~\text{nm}$. The grating couplers are designed to be fully etched type using photonic crystal based metamaterial~\cite{Ding:13}, so that grating couplers, silicon waveguides, and other building blocks can be fabricated with a one-step etching process. The fabricated grating coupler reaches highest coupling efficiency of $-2.4$ dB with 1 dB bandwidth of $39$ nm, as shown in  Fig.~\ref{fig:Devices}(A). Further reductions of coupling loss can be achieved by positioning reflective mirrors under the grating couplers~\cite{Ding:14}.   
High extinction-ratio and low-loss are also critical for the other building blocks, e.g. waveguide crossers, Mach-Zehnder interferometers (MZIs), etc. A network of $16\times16$ crossers is used to swap the signal and idler photons. Waveguide crossers, shown in Fig. \ref{fig:Devices}(B), are designed using the self-imaging effect in a multi-mode interferometer (MMI) and simulated using our in-house code based on three dimensional finite-difference time-domain (FDTD) methods. The insertion loss is measured to be below $-0.1$ dB with crosstalk well below $-40$ dB within C band by using cut-back methods. 
The photons manipulation is based on thermal tuning of the optical phases via heaters. $2\times 2$ MMIs with near $50$:$50$ splitting ratio, tolerant to fabrication errors, are used as beam-splitters in the MZIs. 
Fig.~\ref{fig:Devices}(C) shows the characterization and testing of an exemplary asymmetric MZI. The transmission is measured and normalized to a straight waveguide, and an insertion loss lower than $0.2$ dB is measured, with extinction ratio of more than $35$ dB. The heater is designed to be 100 {\micro\meter} long and 1.8  {\micro\meter} wide with resistance of $~500 {\Omega}$. A $~5 V$ driving voltage results in approximately one free spectral range (FSR) shift of the transmission spectrum, corresponding to a $2\pi$ phase shift. 
In our experiment, we used AMZIs as filters to deterministically separate the generated signal photon with $\lambda_{s}=1539.73$ nm and idler photon with $\lambda_{i}=1549.32$ nm. The AMZI filters used are similar to the one shown in Fig.~\ref{fig:Devices}(C), but with different FSR, here designed as $v_{FSR}=2400$ GHz (or $\lambda_{FSR}=19.2$ nm).  
More results for the MZI characterization are provided in Supplementary section 1.4. 

\setcounter{figure}{0}
\makeatletter 
\renewcommand{\thefigure}{S\@arabic\c@figure}
\makeatother

\setcounter{table}{0}
\makeatletter 
\renewcommand{\thetable}{S\@arabic\c@table}
\makeatother

\begin{figure}[h]
\centering
\includegraphics[width=1.\textwidth]{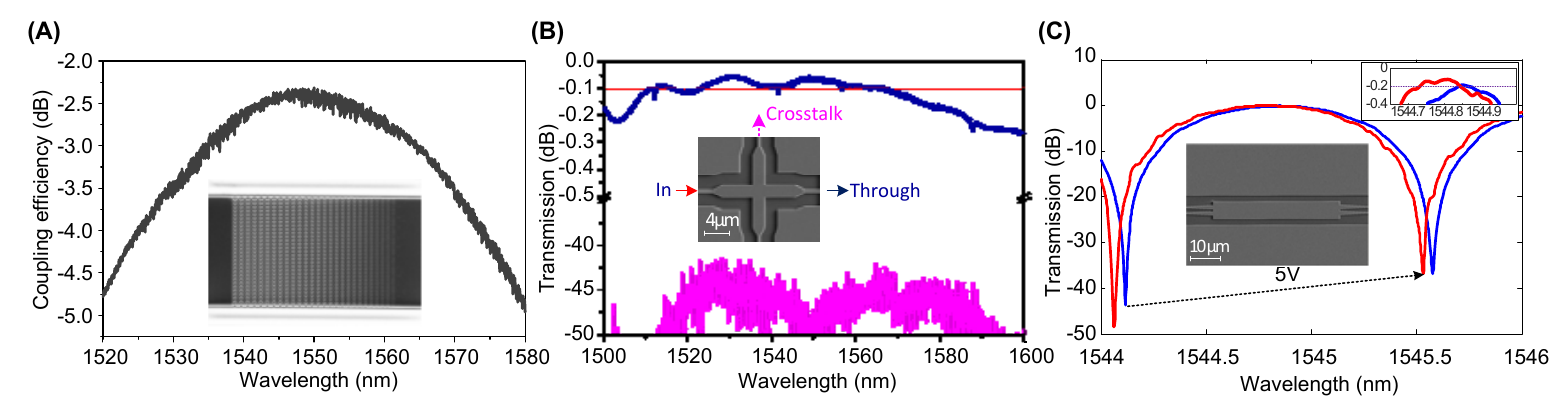}
\caption{Characterizations and SEM (scanning electron microscope) images for (A) optical input/output grating coupler, (B) waveguide crosser, and (C) tunable MZI consisted of two $2\times 2$ MMI beamsplitters and a controllable thermo-optical phase-shifter.}  
\label{fig:Devices}
\end{figure}

\subsection{Experimental setup}
\begin{figure}[t!]
\centering
\includegraphics[width=0.9\textwidth]{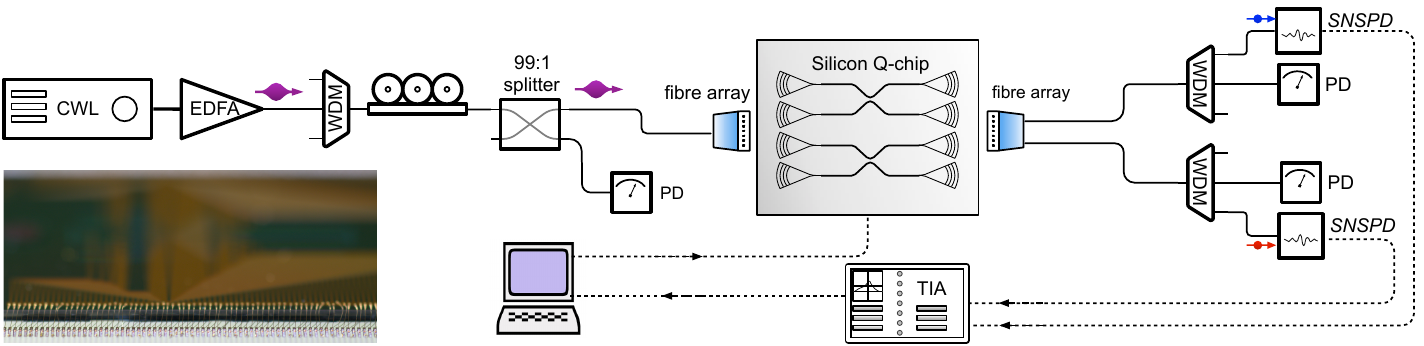}
\caption{Schematic of the experimental setup. 
CWL: continuous-wave laser; EDFA: erbium doped fiber amplifier;  WDM: wavelength-division multiplexer; PD: photo-diode; SNSPD: superconducting nano-wire single-photon detectors; TIA: time interval analyzer. 
Inset: a close view of gold wire bonds on the chip. }  
\label{fig:setup}
\end{figure}

A schematic of the experimental setup is shown in Fig.~\ref{fig:setup}. Most of the components used are off-the shelf telecommunication devices working on C-band wavelengths. 
Non-degenerated photon pairs, with different wavelengths, are generated on-chip via spontaneous four wave mixing (SFWM)~\cite{Silverstone2013}. 
The photon-pair sources are pumped by a continuous-wave (CW) laser source amplified by a commercial erbium doped fiber amplifier (EDFA). Approximately $40$ mw CW light is injected into the chip for photon-pair generation.   
The side bands of the amplified pump are filtered through a wavelength-division multiplexer (WDM) with $\approx 1~\textrm{nm}$ bandwidth and $1.6$ nm channel space. 
The input light polarization is optimized by a polarization controller, then coupled into the chip through grating couplers (Fig. \ref{fig:Devices}(A)). After the chip, light is coupled into optical fibers, then filtered through WDM removing the residual pump photons from the generated single-photons. 
In our experiment, we used pump light with wavelength of $\lambda_p=1544.49$ nm, and generated single photon-pairs with a broadband distribution (approximately $30$ nm bandwidth). 
We post-selected the signal photon at $\lambda_s=1539.73$ nm and idler photon at $\lambda_i=1549.32$ nm, by combing the on-chip AMZI filters and off-chip WDMs. The FSR of the AMZI filters were designed to be $12$ times as that of the WDMs. 
Output CW light is detected by photo-diodes (PD) for device characterizations and for monitoring the coupling and optimizing input polarization. The single-photon pairs are detected through superconducting nano-wire single-photon detectors (SNSPDs) from PhotonSpot\texttrademark, with an average efficiency of $85\%$, $100$ Hz dark counts and $50$ ns dead-time. The SNSPDs output electronic signals are analyzed by a time interval analyzer (TIA). 
The device is wired-bonded on a PCB, as shown in Fig.\ref{fig:setup}. All the 93 thermo-optic phase-shifters can be individually controlled by a computer-interfaced electronic controller, with 96 channels, 12-bits resolution and micro-second speed by Qontrol\texttrademark. Overall all phase-shifters on the chip can be reconfigured at a kHz rate. Each phase-shifter is connected to two separate pads (one for signal and one for ground) to prevent electronic cross-talk. 
A Peltier-cell together with a thermistor and a proportional integrative derivative controller were used to keep the temperature of the photonic device stable. A standard water-cooling system was built to ensure an efficient dissipation of the power injected into the photonic chip. 

\subsection{Linear optical scheme for qudits manipulation}
After the bipartite state $\sum_{k=0}^{d-1} c_k \ket{k}_A\ket{k}_B$ is generated through the coherent excitation of multiple sources and the photon routing, the local manipulation of each qudit is performed using linear-optical schemes. These can in general perform universal operations~\cite{Reck,Oxford:U,Carolan:15}. 
In our device for each photon a triangular structure of MZIs is used, as shown in Fig.\ref{fig:Schematic}A in the main text. 
This scheme can be reconfigured to perform most of the operations relevant to a prepare and measure scenario. For instance, it enables to perform any local projective measurements on the two path-encoded qudits, as described below. 

Projective measurements on the computational basis of a path-encoded photonic qudit are simply obtained by detecting photons in the optical spatial mode associated with the state $|k\rangle$. A projection onto an arbitrary state $|\psi\rangle=\sum_{k=0}^{d-1} a_k |k\rangle$ is obtained by an operation $|\psi\rangle \xrightarrow{\ \hat{U}_d\ } |k\rangle$ followed by a measurement in the computational basis. In general any operation $\hat{U}_d$ can be obtained using only phase-shifters, integrated beamsplitters and MZIs~\cite{Reck,Carolan:15}, whose $2\times 2$ operations are respecively given by 
\begin{equation}
\hat{M}_{PS}(\phi)=
\begin{pmatrix}
  1 & 0 \\
  0 & e^{i\phi} \\
 \end{pmatrix},
 \qquad
 \hat{M}_{BS}=
\frac{1}{\sqrt{2}}\begin{pmatrix}
  i & 1 \\
  1 & i \\
 \end{pmatrix},
 \qquad
 \hat{M}_{MZI}(\theta)=\hat{M}_{BS}\cdot\hat{M}_{PS}(\theta)\cdot\hat{M}_{BS}=
e^{i(\theta+\pi)/2}\begin{pmatrix}
  \sin(\theta/2) & \cos(\theta/2) \\
  \cos(\theta/2) & -\sin(\theta/2) \\
 \end{pmatrix},
 \label{eq:components}
\end{equation}
where the phase is being applied on the second mode interfered (second diagonal element in $\hat{M}_{PS}$). Obtaining the equivalent matrices for the case where the phase-shifter is on the first mode is straightforward. 

The adopted scheme uses the operations in Eq.(\ref{eq:components}) to map $|\psi\rangle \to |k_0\rangle$ for an arbitrary projection $|\psi\rangle$, where photon detection is performed on $|k_0\rangle$. This is done by iterating vector elements elimination in the following way. Up to adding ancillary optical modes, without loss of generality we can consider $d$ to be a power of 2, i.e. $d=2^N$ with $N\geq 1$. In this case, the triangular structure will consist of $N$ layers of MZIs, with the $n$-th layer consisting of $2^{N-n}$ MZIs in parallel (where $n=1,\ldots,N$). Taking $k_0=2^{N-1}$, the values of the phases inside the MZIs $\{ \theta \}$ and in the array of phase-shifters on each mode before the structure $\{ \phi \}$, shown in Fig.~\ref{fig:Schematic}, are chosen such that at each layer the number on non-zero amplitudes in $\{ a_k \}$ is halved. In this way the only non-zero element after the $N$ elimination steps is $a_{k_0}$ associated with the optical mode $k_0$, and therefore the transformation $|\psi\rangle \to |k_0\rangle$ is obtained. Note that this operation is deterministic and no post-selection is required. In our experiment, we implement the correlation measurements by reconfiguring the qudit projectors and collapsing the states into the $|k_0=7\rangle$ computational basis state.  

Pseudo-code for the algorithm used for calculating the phase values is reported in Supplementary Algorithms~\ref{algorithmPhases}--\ref{FullalgorithmPhases}.

\subsection{Characterizations of the sources and projectors}
\begin{figure}[t!]
\centering
\includegraphics[width=0.68\textwidth]{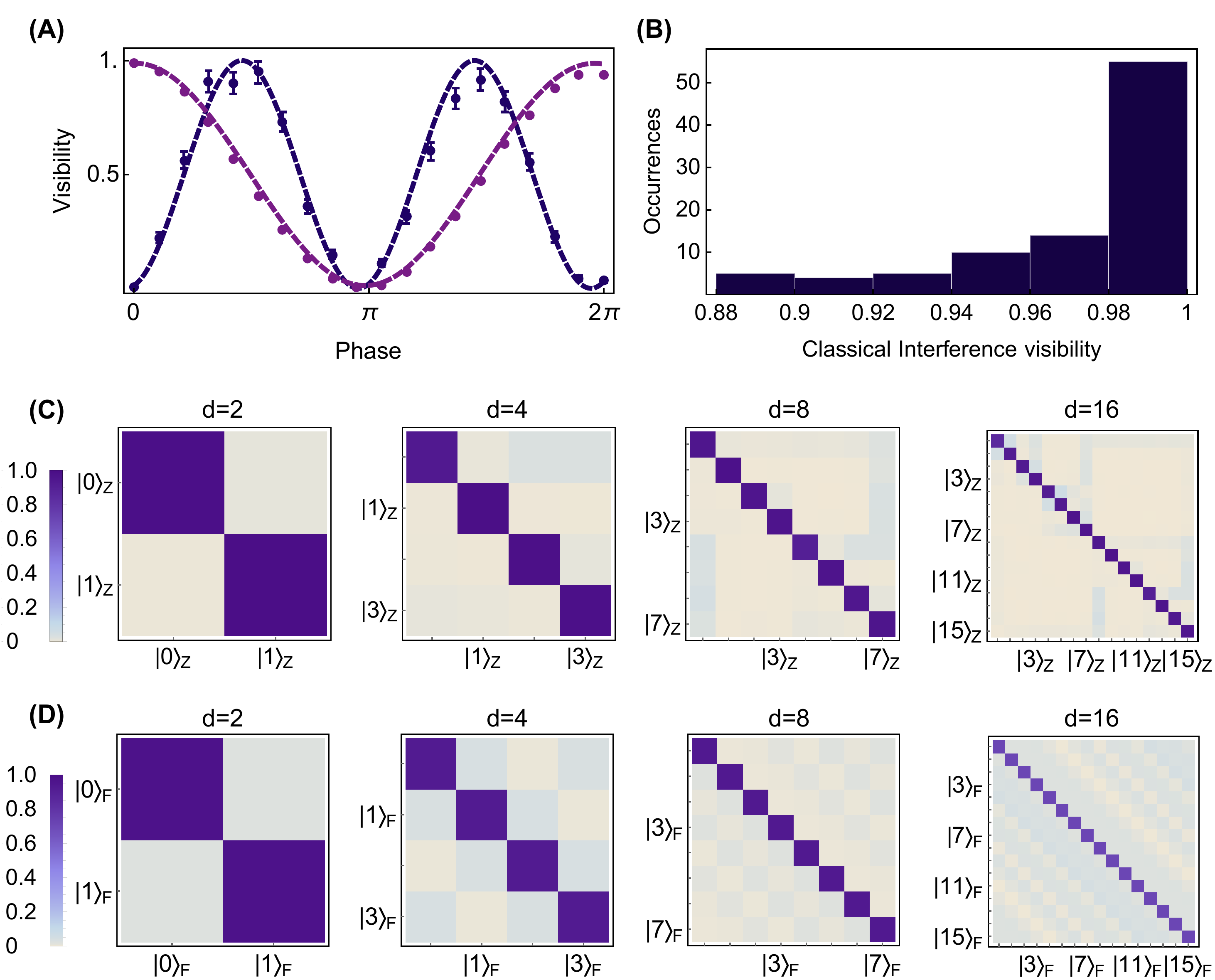}
\caption{Characterization of the photon sources and multidimensional projectors. 
(\textbf{A}) Example of classical interference (purple) and two-photon quantum interference (blue). 
The points are experimental data, while the lines are sinusoidal fittings.  Error bars are given by photon statistics. 
(\textbf{B}) Histogram of the classical interference visibilities for the characterization of all the 93 phase-shifters.  
Measured probabilities distributions preparing and projecting the states on the computational $\hat{Z}$-basis (\textbf{C}) and Fourier $\hat{F}$-basis (\textbf{D}) for qudits with dimension $2$, $4$, $8$ and $16$. }  
\label{fig:characterization}
\end{figure}

We here report the characterization details of the $d$-dimensional qudit projectors and the photon-pair sources. For each thermo-optical phase-shifter, the characterization is obtained by two steps:   
1. measuring the current-voltage (I-V) curve for each heater. A parabolic function was used to fit the I-V curves to include nonlinear response; 
2. mapping electrical power to optical phase. This is obtained by measuring the output optical power as a function of the electrical power dissipated in each phase-shifter. 
The measured classical fringe is fitted to a sinusoid function. 
By repeating these two steps, each phase-shifter can be fully characterized by determining the function of voltage and phase.  
Note that the phase-shifters before the triangular network of MZIs are characterized by reconfiguring the network so that they are embedded inside an MZI, which enables to extract the phase from the interference fringe. 
Figure \ref{fig:characterization}A shows one example of classical interference (purple points). We characterized all the $93$ phase-shifters and obtained high-visibility for each of them. The histogram of the visibilities (defined as $V=\frac{N_{max}-N_{min}}{N_{max}+N_{min}}$) is reported in Fig.~\ref{fig:characterization}B, showing an averaged visibility of $0.973\pm 0.033$. 

After the calibration of the single phase-shifters and MZIs, we continue characterizing the performance of the $d$-dimensional qudit-projectors. We measured the input-output probability distributions for the qudit-projector both on the computational $\hat{Z}$-basis and Fourier $\hat{F}$-basis, defined in the main text.   
The obtained $d\times d$ probabilities distributions are used to quantify the performance of the qudit-projector. 
Fig.~\ref{fig:characterization}C and D reports the measured results for the qudit-projectors. Classical fidelities, defined as $\sum_i \sqrt[]{p_iq_i}$ where $p_i$ and $q_i$ are the measured and ideal output probabilities, were measured, obtaining fidelities of $0.998$, $0.990$, $0.979$ and $0.971$ for $d=2$, $4$, $8$ and $16$ on the $\hat{Z}$-basis; and fidelities of $0.990$, $0.965$, $0.970$ and $0.844$ for $d=2$, $4$, $8$ and $16$ on the $\hat{F}$-basis. All fidelity values for different dimensions are reported in Fig. 1D in main text.  
Note that the results for the $\hat{F}$-basis show higher noise in the qudit operations compared to the $\hat{Z}$-basis, mainly due to residual thermal cross-talks between the phase-shifters ($\hat{Z}$-basis measurements do not matter with the relative phase in each mode). In fact, measurements on the $\hat{F}$-basis required more heaters to be turned on and rely significantly on the phases before the triangular MZI networks, for which we expect higher disturbance. Cross-talk between phase-shifters can be mitigated using optimized designs for the heaters~\cite{Harris} or ad-hoc characterization techniques as in Ref.~\cite{Carolan:15, BPEA}. 

The 16 spiral photon-pair sources were designed identically, with the same length of $1.5$ cm. The SFWM nonlinear effect in silicon is exploited to produce non-degenerate photon-pairs. 
The pump power in each source can be controlled by tailoring the pump distribution across the sources using the input MZIs structure. From each source, around $2$ kHz photon-pairs can be detected by the SNSPD detectors in typical experimental conditions. 
To quantify the indistinguishablity (spectrum overlap) of the $16$ sources, and to measure the uniformity of source brightness, we performed two-photon HOM-like (reversed-HOM) interference experiments for all possible $\binom{16}{2}=120$ pairs of the 16 sources. Fig.\ref{fig:characterization}(A) shows one example of two-photon quantum interference fringe (blue points), with the typical $\lambda/2$ signature compared to the classical fringe. All quantum visibility data are reported in Fig.~1C in the main text.   

\section{Compressed sensing tomographies} 

\begin{figure}[h!]
\centering
\includegraphics[width=0.52\textwidth]{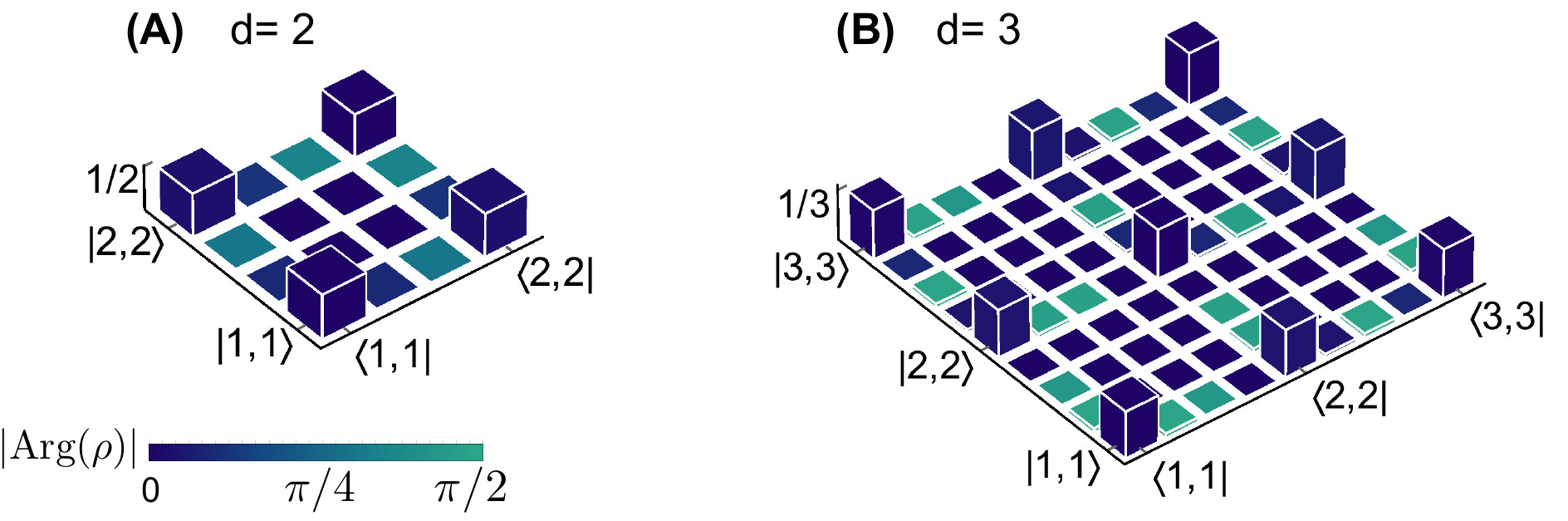}
\caption{
Reconstructed density matrices for the entangled states in dimension (\textbf{A}) 2 and (\textbf{B}) 3. Over-complete tomography techniques are used to reconstruct the states. 
Column heights represent the absolute values $|\rho|$ while colors represent the phases $|\text{Arg}(\rho)|$. The phase information for matrix elements with module $|\rho_{ij}|<0.01$ is approximately randomly distributed and not displayed for clarity. The tomography results for the entangled qudit states in dimension 4, 8 and 12 are reported in Figure 2 in the main text. 
}  
\label{fig:TomoSuppl}
\end{figure}

Quantum state tomography (QST) represents the process of experimentally estimating the full density matrix of a quantum state. While QST represents an invaluable diagnostic tool in the development of quantum technologies, the practical implementation of QST becomes extremely challenging even for the intermediate scale systems that can already be experimentally generated. This is because the density matrix describing the joint state of $n$ subsystems with local dimension $d$ contains $d^{2n}$ entries, which increases rapidly with the system size.
In order to allow QST for the systems involved in existing and near-term quantum technologies, new techniques have been recently developed, with quantum compressed sensing being one of the most prominent ones. \\ 
The idea in compressed sensing, originally developed for classical data analysis, is to exploit the structure that result from data collected in realistic situations to significantly reduce the number of parameters to be determined. Low-rank matrix recovery methods have inspired techniques for the reconstruction of low entropy quantum states, namely compressed sensing quantum state tomography (CSQST), which have been shown to provide speed ups both in term of number of measurement required and post-processing of the data to reconstruct the density matrix~\cite{Gross:2010cv,Gross2011,Flammia2012}. These methods have been implemented to reconstruct density matrices of systems composed of up to seven trapped-ion qubits~\cite{CompressTomo}.
While CSQST was originally developed for low entropy states, it is also believed to be appropriate in the cases where the density matrix is not low rank~\cite{CompressTomo}. In our experiment the low entropy of the quantum states generated is also certified by the high quality of the quantum correlations measured in the other protocols studied here, which further justifies the use of CSQST.\\
The experimental procedure for CSQST requires projective measurements in the eigenbases of $m$ operators randomly sampled from a chosen ortho-normal basis $\{\omega_i \}_{i=1}^{d^{2n}}$ of the operators space. Indicating with a vector $\boldsymbol{y}$ the measured statistics for all the outcomes obtained measuring on these bases, and with $\mathcal{A}(\mathcal{X})$ the list of expectation values of the outcomes for a generic input $\mathcal{X}$, where $\mathcal{A}$ is a linear operator formalising the measurement process~\cite{Flammia2012}, the estimated matrix is given by the solution of the semi-definite program (SDP)
\begin{align}
\tilde{\rho}= &\text{argmin } \text{Tr}\mathcal{X} \\ 
&\text{s.t.} \quad \mathcal{X} \geq 0,\  \|\mathcal{A}(\mathcal{X}) -  \boldsymbol{y} \|_2^2  \leq \epsilon, \nonumber
\end{align}
and finally renormalising $\rho=\tilde{\rho}/\text{Tr}(\tilde{\rho})$. This is a convex trace norm minimization constrained on semi-positive definite matrices, for which efficient convex optimization programs have been extensively studied and are readily available~\cite{yalmip,sedumi}. Here $\epsilon$ is a control parameter related to the amount of noise in the data, and for which we adopt the same heuristics used in Ref.~\cite{CompressTomo}. Theoretical results have shown that in order to reconstruct a $N\times N$ density matrix of rank $r$, the measurements of $O(Nr \log(N^2))$ operators is sufficient, in contrast with $O(N^2)$ used for standard approaches~\cite{Gross:2010cv,Gross2011,Flammia2012}, obtaining approximately a quadratic advantage. Moreover, this scaling is valid for any basis $\{\omega_i \}$~\cite{Gross2011}. 
In our experiment, a total of 50, 122 and 228 operators are measured to reconstruct the state in dimension 4, 8, and 12, respectively, following reported measurement procedures~\cite{CompressTomo,Bolduc:2016fv}. The SDP for the density matrix reconstruction was performed using the SeDuMi solver~\cite{sedumi} on a standard laptop, and took few seconds for $d=4$, approximately five minutes for $d=8$ and approximately one hour for $d=12$. The reconstructed density matrices obtained are shown in Fig.2 in the main text, and present fidelities  $96\%$, $87\%$ and $81\%$ with the $d$-dimensional bipartite maximally entangled state associated. For systems with lower local dimensions, that is $d=2$ and $d=3$, the density matrices were reconstructed using standard over-complete tomography techniques, and are shown in Fig.~\ref{fig:TomoSuppl}. The obtained fidelities are $99\%$ and $97\%$, respectively.

\section{Dimension witness}

For the certification of the local dimension of the quantum states generated we follow the approach of Ref.~\cite{Sikora:DIDW}. In particular the bounds are obtained using the following theorem:

\begin{theorem}[\cite{Sikora:DIDW}] 
Suppose a correlation $\vec{p} = p(ab|xy)$ is produced by two separated parties Alice and Bob who do not have access to shared randomness but possess a shared quantum state $\ket{\psi}$ on $\C^{d_1} \otimes C^{d_2}$.
Then
\begin{equation*}
	\lceil \mathcal{D}(\vec{p})\rceil=\left\lceil\frac{1}{f(\vec{p})}\right\rceil \le \min\set{d_1,d_2}
\end{equation*}
where
\begin{equation}
	f(\vec{p}) := \min_{y,y'} \sum_{b,b'} 
			\min_x
			\left( \sum_a \sqrt{p(ab|xy) p(ab'|xy')} \right)^2.
\label{eq:Bound}
\end{equation}
\label{thm:Bound}
\end{theorem}

To give an upper bound to $f(\mathbf{p})$, and consequently a lower bound to the local dimension, instead of minimizing over all possible values $y,y'$, and $x$ in Eq.~\ref{eq:Bound} we can perform measurements for some fixed measurement settings, $y,y'$ for Bob and $x$ for Alice, and evaluate the function
\begin{equation*}
	f_{y,y',x}(\vec{p}) := \sum_{b,b'} 
			\left( \sum_a \sqrt{p(ab|xy) p(ab'|xy')} \right)^2.
\end{equation*}
Indeed, for any choice of settings, $y,y'$ and $x$ the value $f_{y,y',x}(\vec{p})$ upper bounds the optimized value $f(\vec{p})$. Therefore, the inverse, $1/f_{y,y',x}(\vec{p})$, lower bounds the local dimension $\min\set{d_1,d_2}$.
In the cases we consider, we show below that we can choose $y,y',x$ so that $f_{y,y',x}(\vec{p}) = f(\vec{p})$. Therefore, instead of collecting the full correlation data for all possible measurement settings, it will suffice to fix the measurement setting for Alice and collect the data for two possible measurement settings for Bob. Note that, irrespective of the correlation $\vec{p}$, it will always be possible to only gather measurement statistics for two different measurement settings for Bob by simply picking the two values of $y$ and $y'$ which attain the minimum in Equation~\ref{eq:Bound}. In contrast, we might need to use many different measurement settings for Alice in order to obtain the  the best bound. This is because Alice's setting $x$ attaining the minimum in Equation~\ref{eq:Bound} could depend on $b$ and $b'$.

\paragraph{Magic Square correlations.} In the Magic Square game \cite{Mermin1990,PeresMS,AravindMS} each of Alice and Bob have three different
measurement settings: $\set{1,2,3}$. Alice's outputs are $A = \set{a \in\set{+1,-1}^3 : a_1 a_2 a_3 =1 }$ and Bob's
outputs are $B =\set{b \in \set{+1,-1}^3 : b_1 b_2  b_3 = -1}$. The players win the game, if their outputs, $a$ and $b$,
satisfy $a_y = b_x$, where $x$ and $y$ are the measurement settings. Although Magic Square game cannot be won using classical correlations, the following quantum correlation $\vec{p}$ allows to win the Magic Square game irrespective of the selected measurement settings:
\begin{equation}
	\vec{p} = p(ab|xy) = 
	\begin{cases}
		\frac{1}{8} & \text{if $a_y = b_x$} \\
		0			& \text{otherwise}
	\end{cases}
\label{eq:MS}
\end{equation}
This correlation can be realized by measurements on a bipartite maximally entangled state of local dimension 4.
In particular, the $i$th measurement setting for Bob (Alice) corresponds to measuring in the orthonormal basis diagonalizing all the operators in the $i$th column (row) in the square grid shown in Fig.~3B in the main text, where the 4-dimensional local Hilbert space is represented as a 2-qubit space. 
The outcome Bob (Alice) obtains corresponds to a simultaneous eigenvector of the the three operators in the respective column (row)  with
eigenvalues $\lambda_1, \lambda_2, \lambda_3 \in \set{+1,-1}$. Since the along any row (column) the operators multiply to $I$ ($-I$). The eigenvalues obtained by Alice will multiply to $+1$, while those obtained by Alice will multiply to $-1$. Therefore, the eigenvalue strings $(\lambda_1,\lambda_2,\lambda_3)$ obtained by each of the players correspond to valid answers.

Using $y = x = 1$ and $y'=2$
we find that 
\begin{equation}
	f_{y,y',x}(\vec{p}) = \sum_{b,b' \in B} 
			\left( \sum_{a\in A} \sqrt{p(ab|11) p(ab'|12)} \right)^2
            = \sum_{b,b' \in B}  \left(\sqrt{\frac{1}{8} \cdot \frac{1}{8}} \right)^2
            = \frac{\lvert B \rvert^2}{64} = \frac{16}{64} = \frac{1}{4}
\label{eq:MSquare}            
\end{equation}
and therefore we can certify that the local dimension of the measured system is at least four. 
Note that we can reach these conclusions solely from observing correlation $\vec{p}$ and without making any assumption about the performed measurements or the shared state, though we do need the assumption that shared randomness is not free. The above analysis applies to the ideal case with no experimental error. In an actual experiment we can gather outcome statistics to obtain probabilities, $\hat{p}(ab|xy)$. If the experiment has low noise, the correlation $\hat{\vec{p}}$ will be very close to $\vec{p}$ and we will get $\lceil 1/f_{y=1,y'=2,x=1}(\hat{\vec{p}}) \rceil = 4$ thus certifying that the dimension of the measured system is 4. In particular, in our experiment we obtained correlation, $\vec{p}^\text{I}_4$, with $1/f_{y=1,y'=2,x=1}(\vec{p}^\text{I}_4) = 3.565$ (see Table~\ref{tableDW}) hence indeed certifying dimension 4. As discussed above, in order to evaluate $f_{y=1,y'=2,x=1}$ we only need to gather measurement statistics for measurement settings $y=1$ and $y'=2$ for Bob and $x=1$ for Alice. According to Fig.~\ref{fig:DW}B, this amounts to measuring both of Alice's qubits in the ${Z}$-basis ($x=1$) and either measuring Bob's first qubit in the ${X}$-basis and the second one in the ${Z}$-basis ($y=1$) or the other way around ($y'=2$), where $X, Y$, and $Z$ are the Pauli matrices. We refer to this as ``scenario I'' and denote the experimentally observed distribution by $\vec{p}_4^\text{I}$.

We can also think of playing two copies of the Magic Square game in parallel, where the players win if the winning conditions for both games are satisfied simultaneously. The correlation $\vec{p_2}$ that lets us win this game can be realized by measurements on a bipartite maximally entangled state of local dimension 16, where each player holds the equivalent of a 4-qubit system. Via a calculation similar to the one in \ref{eq:MSquare}, we find that $f_{y=(1,1),y'=(2,2),x=(1,1)}(\vec{p_2}) = 1/16$, and hence correlation $\mathcal{D}(\vec{p}) \ge 16$ and this correlation certifies dimension 16.

\paragraph{Magic Pentagram correlations.} In the Magic Pentagram game~\cite{Mermin1990} each of Alice and Bob have five different
measurement settings: $\set{1,2,3,4,5}$ and we think of each of these settings as identifying one the five lines in the pentagram in Fig.~3B in the main text. 
Alice's and Bob's outputs are $C^+ = \set{c\in\set{+1,-1}^4: c_1c_2c_3c_4 = 1}$ for settings 2 to 4 and their outputs are $C^- =\set{c\in\set{+1,-1}^4: c_1c_2c_3c_4 = -1}$ for setting 5. We think of the $i$th output position as assigning value to the $i$th point on the line corresponding to the measurement setting. For measurement settings $x$ and $y$, the players win the game, if their outputs, $a$ and $b$, assign the same value to all the points belonging to both lines $L_x$ and $L_y$.

For instance if we ask Bob the fifth line ($y=5$) and Alice the third one ($x=3$) then these two lines intersect in precisely one point (the 1st point on $L_5$ and the 5th point on $L_3$). So to win, the first position in Bob's answer string should equal the last position in Alice's answer string; for example, $b=({+1},{-1},{+1},{+1})$ and $a =({+1},{-1},{-1},{+1})$ would be a winning answer pair.

Just like in the case of Magic Square, there exists a quantum correlation $\vec{p}$ that can be used to win the Magic Pentagram game for all measurement settings but no classical correlation would achieve this. Specifically,
\begin{equation}
	\vec{p} = p(ab|xy) = 
	\begin{cases}
		\frac{1}{32} & \text{if $a$ and $b$ satisfy conditions (1) and (2) above} \\
		0			& \text{otherwise}
	\end{cases}
\label{eq:MSPent}
\end{equation}
and it can be realized by measurements on a maximally entangled state of local dimension 8, locally equivalent to a 3-qubit space.
In particular, the $i$th measurement setting for Bob (Alice) corresponds to measuring in the orthonormal basis diagonalizing all the commuting observables on the $i$th line See Fig.3. The outcome corresponds to a simultaneous eigenvector of all four operators and the players respond with a string of eigenvalues $(\lambda_1,\dotsc,\lambda_4)$ that correspond to their measurement outcome. This string will be a valid answer since the observables on the first line multiply to $-I$ and to $I$ along all the other lines.

Fixing $y = x = 3$ and $y' = 5$, we find that 
\begin{equation}
	f_{y,y',x}(\vec{p}) = \sum_{b\in C^+, b'\in C^-} 
			\left( \sum_{a\in C^+} \sqrt{p(ab|33) p(ab'|35)} \right)^2
            = \sum_{b\in C^+, b'\in C^-} p(bb|33) p(bb'|35)
            = \sum_{b\in C^+} \frac{1}{32} \sum_{b'\in C^-} p(bb'|35)
            = \frac{\lvert C^+ \rvert \cdot 8}{32\cdot 32 } = \frac{1}{8}
\end{equation}
as $p(ab|33) = \delta_{ab}/32$ and $p(bb'|35) = \delta_{b_5b'_1}/32$. It now follows that we can certify that the local dimension of the measured system is at least~8 by observing correlation $\vec{p}$.

\paragraph{Scenario I.} The measurements we perform in this scenario are based on the optimal strategies for Magic Square and Magic Pentagram games that we explained above. For dimensions $d=4,8$ we prepare the $d$-dimensional maximally entangled state $\ket{\psi_d^+} = \frac{1}{\sqrt{d}} \sum_{k=0}^{d-1}\ket{k,k}$ and then perform the measurements corresponding to optimal strategies for Magic Square game ($d=4$) or Magic Pentagram game ($d=8$). Instead of gathering statistics for all possible measurement settings in these games we choose only one setting for Alice (setting 1 for Magic Square and setting 3 for Magic Pentagram) and two different settings for Bob (settings 1, 2 for Magic Square and settings 3, 5 for Magic Pentagram). We argued above that in the absence of experimental error, these correlations would enable us to certify dimensions 4 and 8 respectively. Experimentally we obtain correlations $\vec{p}^\text{I}_4$ and $\vec{p}^\text{I}_8$ yielding values $\lceil \mathcal{D}(\vec{p}^\text{I}_4) \rceil = 4$ and $\lceil \mathcal{D}(\vec{p}^\text{I}_8) \rceil = 8$ thus certifying dimensions 4 and 8 respectively.

In order to certify intermediate dimensions $4 < d < 8$ we could prepare state
\begin{equation}
	\ket{\hat{\psi}_d^+} = \frac{1}{\sqrt{d}} \sum_{k=0}^{d-1}\ket{k,k}
\end{equation}
where $\ket{k}\in \mathbb{C}^8$. Note that although essentially  $\ket{\hat{\psi}_d^+}$ is the $d$-dimensional maximally entangled state, it is embedded in dimension $\mathbb{C}^8 \otimes \mathbb{C}^8$ so we can measure it using the $8$-dimensional basis measurements from the optimal strategy of Magic Pentagram game. As before, we only use measurement setting 3  for Alice and settings 3 and 5 for Bob. We only performed experiment for even dimensions and the experimentally observed correlation $\vec{p}^\text{I}_6$ yielded $\lceil \mathcal{D}(\vec{p}^\text{I}_6) \rceil = 6$ successfully certified dimension 6 (see Table~\ref{tableDW} ).

In order to certify even dimensions $8 < d < 16$ we use a similar approach but perform measurements from the optimal strategy for two copies of Magic Square instead of Magic Pentagram game. Specifically, we prepare state
$\ket{\hat{\psi}_d^+} = \frac{1}{\sqrt{d}} \sum_{k=0}^{d-1}\ket{k,k}$
where $\ket{k}\in \mathbb{C}^{16}$ and measure it using measurement setting $(1,1)$ for Alice and settings $(1,1)$ and $(2,2)$ for Bob.  For Alice this amounts to measuring her state in the ${Z}$-basis while Bob either measures his first and third qubit in the ${X}$-basis and the remaining two in the ${Z}$-basis (setting (1,1)) or the other way around (setting (2,2)). From Table~\ref{tableDW} we see that the experimentally obtained correlations $\vec{p}^\text{I}_d$ were capable of certifying the true dimension for all $d<12$.

\paragraph{Scenario II.} In this scenario, to obtain the correlation $\vec{p}_d^\text{II}$, we prepare the $d$-dimensional maximally entangled state $\ket{\psi^+_d}$ for $4 \le d \le 15$ and measure each of the two local systems in the $\hat{Z}$-basis. In the ideal case, we would obtain a correlation $p_d(ab|11) = \delta_{ab}/d$, where $a,b \in \set{0,\dotsc,d-1}$. Although a similar conclusion could also be obtained in a more straightforward manner, we can compute the bound $\mathcal{D}(\vec{p}_d)$ to see which dimension would be certified by experimentally observing this correlation. Since there is only a single measurement setting for both Alice and Bob (\textit{i.e.}, a single choice for the values $y,y'$ and $x$), we see that
\begin{equation}
	f(\vec{p}_d) 
			 = \sum_{b,b'}  \left( \sum_a \sqrt{p(ab|11) p(ab'|11)} \right)^2
			 = \sum_{b,b'}  \left( \sum_a \frac{\delta_{ab}}{\sqrt{d}} \sqrt{ p(ab'|11)} \right)^2
			 = \frac{1}{d} \sum_{b,b'}  p(bb'|11) = \frac{1}{d}
\end{equation}
as the only non-zero term in the sum over $a$ is the one where $a=b$. Therefore, it follows that $\mathcal{D}(\vec{p}_d) = 1/f(\vec{p}_d) = d$ and the correlation $\vec{p}_d$ can be used to certify dimension $d$. Of course, experimentally we do not observe exactly $\vec{p_d}$. Nevertheless, the experimentally observed correlations $\vec{p}^\text{II}$ resulted in  $\lceil \mathcal{D}(\vec{p}_d^\text{II}) \rceil = d$ for all $4 \le d \le 14$ (see Table~\ref{tableDW}).

\begin{table}[ht!]
\centering
\begin{tabular}{c| c c c c c c c } 
 \hline \hline \noalign{\smallskip} 
Dim & 4 & 6 & 8 & 10 & 12 & 14 & 15 \\
\hline \noalign{\smallskip}
Scenario I & $3.565\pm 0.021$ & $5.643 \pm 0.033$ & $7.221 \pm 0.054 $ & $9.147 \pm 0.032 $  & $10.791 \pm 0.042 $  & $12.076 \pm 0.041 $  & $12.573 \pm 0.046 $
\\ 
Scenario II & $3.901\pm 0.005$ & $5.755 \pm 0.017$ & $7.553 \pm 0.022 $ & $9.499 \pm 0.026 $  & $11.382 \pm 0.029 $  & $13.056 \pm 0.036 $  & $13.975 \pm 0.036 $ 
\\ \hline  
\end{tabular}
\caption{Experimentally certified local dimension of the generated entangled qudit states. 
The lower bounds on the dimension, and are measured by adopting the two different measurement scenarios I and II, respectively. These data are used to plot Fig. 3 in the main text. Error bars are given by photon Poissonian noise.}
\label{tableDW}
\end{table}

\section{Multidimensional Bell inequality}

\subsection{Bell inequalities for two-qudit maximally entangled states.}
Here we present in more details the class of Bell inequalities we used to test non-locality of the maximally entangled states of two qudits, which were recently derived in \cite{SATWAP}. While we will focus on the particular case of two measurements per observer, it is worth noticing that the class of Bell inequalities found in \cite{SATWAP} allows for any number of measurement settings. Let us consider a Bell scenario in which two spatially separated parties $A$ and $B$ share some quantum state $\hat{\rho}$. We then assume that on their shares of this state each party performs one of two measurements,\footnote{By a quantum measurement we denote a set $\{\hat{M}_{a|x}\}_a$ of positive semi-definite operators, $\hat{M}_{a|x}\succeq 0$ for any $a$, forming a resolution of the identity $\mathbbm{1}$ acting on the corresponding Hilbert space, that is,
$\sum_{a}\hat{M}_{a|x}=\mathbbm{1}$.} 
$A_x=\{\hat{M}_{a|x}\}_a$ for $A$ and $B_y=\{\hat{M}_{b|y}\}_b$ for $B$ with $x,y=1,2$. Each measurement yields one of $d$ possible outcomes, which are labelled by $0,\ldots,d-1$. 

Such local measurements, after many repetitions, lead to correlations that are described by 
a collection of joint probabilities
\begin{equation}\label{correlations}
\{p(ab|xy)\}_{a,b \in \{0,\ldots,d-1\}}
\end{equation}
with $x,y=1,2$, where each $p(ab|xy)\equiv p(A_x=a,B_y=b)$ is the probability that $A$ and $B$ obtain outcomes $a$ and $b$ upon performing the measurements $A_x$ and $B_y$, respectively, and can be expressed by the Born rule as
%
$p(ab|xy)=\mathrm{Tr}[\hat{\rho}(\hat{M}_{a|x}\otimes \hat{M}_{b|y})]$ (see Fig.~1E in the main text). 

The class of Bell inequalities used to test non-locality of the maximally entangled states is
\cite{SATWAP}:
\begin{equation}\label{Bellproba_app}
I_{d}:=\sum_{k=0}^{\left \lfloor d/2\right \rfloor-1}\left(
\alpha_k \mathbbm{P}_k -\beta_k \mathbbm{Q}_k \right )\leq C_d,
\end{equation}
where the expressions $\mathbbm{P}_k$ and $\mathbbm{Q}_k$ are defined as
\begin{equation}
\mathbbm{P}_k=P(A_1=B_1+k)+P(B_1=A_{2}+k)+P(A_2=B_2+k)+P(B_2=A_{1}+k+1)
\end{equation}
and
\begin{equation}
\mathbbm{Q}_k=P(A_1=B_1-k-1)+P(B_1=A_{2}-k-1)+P(A_2=B_2-k-1)+P(B_2=A_{1}-k).
\end{equation}
The coefficients $\alpha_k$ and $\beta_k$ are given by
\begin{equation}\label{alfabeta}
\alpha_k=\frac{1}{2d}\left[g(k)+(-1)^d\tan\left(\frac{\pi}{4d}\right)\right],\qquad
\beta_k=\frac{1}{2d}\left[g\left(k + 1/2 \right)-(-1)^d\tan\left(\frac{\pi}{4d}\right)\right],\qquad
\end{equation}
with $g(k)=\cot[\pi(k+1/4)/d]$. Here, $P(A_x=B_y+k)$ is the probability that upon measuring $A_x$ and $B_y$, the results obtained by Alice and Bob differ by $k\!\!\mod d$, that is
\begin{equation}
P(A_x=B_y+k)=\sum_{a=0}^{d-1}p((a+k \!\!\!\!\mod d)a|xy).
\end{equation} 
Notice that by replacing the above $\alpha_k,\beta_k$
with $\alpha_k=\beta_k=1-2k/(d-1)$ one obtains the CGLMP Bell expression \cite{CGLMP}.

For our convenience let us now rewrite the inequality (\ref{Bellproba_app})
in the correlator form. As here we work with more outcomes than two, we need to 
appeal to the notion of generalized complex correlators, which are defined 
as the two-dimensional Fourier transform of the joint distributions (\ref{correlations}),
\begin{equation}\label{Fourier}
\langle A_x^k B_y^l\rangle=\sum_{a,b=0}^{d-1}\omega^{ak+bl}p(ab|xy),
\end{equation}
for $x,y=1,2$ and $k,l=0,\ldots,d-1$. We denote the root of unity as $\omega$, with $\omega = \text{exp}(2\pi i /d)$. Notice that when $d =2$, the definition of the correlators $\langle A_x B_y \rangle = P(A_x = B_y) - P(A_x \neq B_y)$ naturally also corresponds to taking the discrete Fourier transform.  Expressing the probabilities $p(ab|xy)$  in terms of the generalized correlators and then plugging into (\ref{Bellproba_app}), we can bring the latter to 
\begin{eqnarray}\label{BellIneq1}
\widetilde{I}_d&\!\!\!:=\!\!\!&\sum_{l=1}^{d-1}\left[a_l\langle A_1^lB_1^{d-l}\rangle+a_l^{*}\omega^l\langle A_1^l B_2^{d-l}\rangle+a_l\langle A_2^l B_2^{d-l}\rangle+a_l^{*}\langle A_2^l B_1^{d-l}\rangle\right]\nonumber\\
&\!\!\!=\!\!\!&2\sum_{l=1}^{\lfloor d/2\rfloor-1}\mathrm{Re}\left[\langle A_1^l\bar{B}_1^l\rangle+\langle A_2^l\bar{B}_2^l\rangle\right]
\leq \mathcal{C}_d,
\end{eqnarray}
where $\mathcal{C}_d$ is its classical bound given by 
\begin{equation}
\mathcal{C}_d=\frac{1}{2}\left[ 3 \cot\left(\frac{\pi}{4d}\right) - \cot\left(\frac{3\pi}{4d}\right)\right] - 2,
\end{equation}
and $\bar{B}_i^{l}$ are new variables defined as $\bar{B}_1^l=a_lB_{1}^{d-l}+a^{*}_l\omega^lB_{2}^{d-l}$ and $\bar{B}_2^{l}=a_lB_2^{d-l}+a_l^{*}B_{1}^{d-l}$ with $a_l=\omega^{(2l-d)/8}/\sqrt{2}$. Notice that in the quantum case $A_x$ and $B_y$
are unitary operators of eigenvalues $1,\omega,\ldots,\omega^{d-1}$, whereas $A_x^{k}$ and $B_y^{l}$ are simply their matrix powers. Then, the complex correlators can be easily expressed with the aid of the Born rule as 
$\langle A_x^k B_y^l\rangle=\Tr[\hat{\rho}(A_x^k\otimes B_y^l)]$. We thus can think of $A_x$
and $B_y$ as quantum observables whose results are for our convenience labelled by 
the roots of unity $1,\omega,\ldots,\omega^{d-1}$.

The maximal quantum violation of the inequality (\ref{BellIneq1})
is $\mathcal{Q}_d=2(d-1)$ and it is achieved with the maximally entangled state of two qudits and the following observables 
\begin{equation}\label{CGLMPmeasurements}
A_x = U_x^{\dagger}F\Omega F^{\dagger}U_x,\qquad 
B_y = V_yF^{\dagger}\Omega F V_y^{\dagger}
\end{equation}
where $\Omega=\mathrm{diag}[1,\omega,\ldots,\omega^{d-1}]$, $F$
is the Fourier matrix given by 
\begin{equation}
F = \frac{1}{\sqrt{d}}\sum_{i,j=0}^{d-1}\omega^{ij}\ket{i}\!\bra{j},
\end{equation}
and, finally, $U_x$ and $V_y$ are unitary rotations defined as
\begin{equation}
U_x = \sum_{j=0}^{d-1}\omega^{j\theta_x}\proj{j},\qquad 
V_y = \sum_{j=0}^{d-1}\omega^{j\zeta_y}\proj{j}
\end{equation}
with $\theta_1 = 1/4$, $\theta_2 = 3/4$, and $\zeta_1 = 1/2$ and $\zeta_2 = 1$.
Notice that the eigenprojectors of Alice's and Bob's measurements $A_x$ and $B_y$ are given by 
\begin{equation}\label{ax}
\ket{a}_x=\frac{1}{\sqrt{d}}\sum_{k=0}^{d-1}\mathrm{exp}\left[i2\pi k(a-\theta_x)/d\right] \ket{k}
\end{equation}
and
\begin{equation}\label{by}
\ket{b}_y=\frac{1}{\sqrt{d}}\sum_{k=0}^{d-1}\mathrm{exp}\left[i2\pi k(-b+\zeta_y)/d\right] \ket{k},	
\end{equation}
respectively, where $a,b \in \{0,\ldots,d-1\}$.

\subsection{Bell inequalities for partially entangled states of two-qutrits} 

We now investigate whether the inequalities (\ref{BellIneq1}) can be generalized to a class of Bell inequalities that are maximally violated by some target partially entangled states. Note that the CGLMP Bell inequality is also of the form (\ref{BellIneq1}) and is maximally violated by a partially entangled state, with the corresponding Bell expression also given by (\ref{Bellproba_app}) but with different coefficients than 
(\ref{alfabeta}). Therefore a natural choice of Bell inequalities to study are those of the form (\ref{BellIneq1}). 
For simplicity let us consider the case of $d=3$, for which 
(\ref{Bellproba_app}) gives the following class of Bell inequalities involving two parameters
%
$\alpha_0\mathbbm{P}_0-\beta_0\mathbbm{Q}_0\leq C$,
%
however, we can always divide the whole expression by one of them, say $\alpha_0$ (provided that it is positive), reducing the number of free parameters to one.
As a result we obtain the following class of Bell inequalities
\begin{eqnarray}\label{class}
I_3(\xi)&\!\!\!:=\!\!\!&P(A_1=B_1)+P(A_2=B_2)+P(A_1=B_2-1)+P(A_2=B_1)\nonumber\\
&&-\xi[P(A_1=B_1-1)+P(A_2=B_2-1)+P(A_1=B_2)+P(A_2=B_1+1)]\leq C_3(\xi),\nonumber\\
\end{eqnarray}
parametrized by a single parameter which we denote $\xi$ and which is defined in terms of the parameters (\ref{alfabeta}) as  $\xi = \beta_0/\alpha_0$. It turns out
that the classical bound of these inequalities can be easily found by looking for local deterministic strategies that maximize $I_3(\xi$) thus yielding its bound within local hidden variable theories, which is 
\begin{equation}
C_3(\xi)=\left\{
\begin{array}{ll}
4\xi, & \mathrm{if}\,\,\, \xi\leq -1,\\
3+\xi, & \mathrm{if}\,\,\, -1\leq \xi\leq 1,\\
2, & \mathrm{if}\,\,\, \xi\geq 1.
\end{array}
\right.
\end{equation}
Moreover, numerical tests using the Navascu\'es-Pironio-Ac\'in (NPA) hierarchy \cite{npa2007}
indicate that for $\xi \leq -1$, the Bell inequality (\ref{class}) is trivial, meaning that
its maximal quantum violation equals its classical bound. Consequently, in what follows
we will concentrate on the case $\xi>-1$. It is then not difficult to see that for $\xi=1$ the class (\ref{class}) reproduces the well-known CGLMP Bell inequality, which is known to be maximally violated by the partially entangled state \cite{Durt2002}
\begin{equation}\label{partially}
\ket{\psi_\gamma}=\frac{1}{\sqrt{2+\gamma^2}}(\ket{00}+\gamma\ket{11}+\ket{22})
\end{equation}
with $\gamma=(\sqrt{11}-\sqrt{3})/2$, whereas for $\xi=(\sqrt{3}-1)/2$ it gives the Bell inequality maximally violated by the maximally entangled states presented in the correlator form in (\ref{BellIneq1}). In both cases the observables (\ref{CGLMPmeasurements}) are used.

The question we want to answer now is whether by changing $\xi$ between the above two values we can obtain Bell inequalities maximally violated by partially entangled states (\ref{partially}) for various values of $\gamma$. To answer this question let us first take
the optimal CGLMP measurements (\ref{CGLMPmeasurements}) and compute the value of the Bell expression for the state (\ref{class}). This gives us the following function of $\xi$ and $\gamma$:
\begin{equation}\label{expression}
\mathcal{I}(\xi,\gamma)=\frac{4[3+\gamma(2\sqrt{3}+\gamma-\xi\gamma)]}{3(2+\gamma^2)}.
\end{equation}
To find its maximal value for a fixed $\xi$, we need to satisfy the following condition
$\partial \mathcal{I}(\xi,\gamma)/\partial \gamma=0$. Solving this equation is equivalent to finding the root of a second degree polynomial in $\gamma$, and the extremum is found to be
%
$\gamma_{+}(\xi)=[(4\xi^2+4\xi+25)^{1/2}-2\xi-1]/2\sqrt{3}$,
%
for which the maximal value of (\ref{expression}) for a fixed $\xi$ is 
%
%
\begin{equation}\label{expression2}
\mathcal{I}_{\max}(\xi)=\frac{1}{3}\left[5 - 2 \xi + \sqrt{25 + 4 (\xi+1) \xi}\right].
\end{equation}

Of course, the above derivation is not a proof that this is the maximal quantum violation of the Bell inequality (\ref{class}), however, based on our numerical study we conjecture this to be the case. Notice first that for $\xi=1$ and $\xi=(\sqrt{3}-1)/2$, the expression
(\ref{expression2}) reproduces the maximal quantum violations of the CGLMP Bell inequality and the one in (\ref{BellIneq1}). Then, we have tested our conjecture for other values of $\xi$ by using the NPA  hierarchy, which we implemented using the Yalmip toolbox \cite{yalmip} and the SeDuMi solver \cite{sedumi} in Matlab. The NPA hierarchy provides outer approximations to the quantum set of correlations, and for a given Bell inequality, it allows one to find an upper bound on the maximal quantum violation of the Bell esxpression.  We employed this technique for values of $\xi\in[-0.99,100]$ with the step 0.01, and for all these values of $\xi$ the value obtained agrees with (\ref{expression2}) up to solver precision $~10^{-8}$, which is a strong implication that it is the maximal quantum violation of the corresponding inequality. Note that for $\xi\in[-0.99,-42]$, the level $1+AB$ of the hierarchy was sufficient, while for $\xi\in[42,100]$ we used the level $2$, except for a small amount of values in the interval $[85,100]$ for which the level $2 + AAB$ was necessary.

Let us finally write the class of inequalities (\ref{class}) in 
the correlator form 
\begin{eqnarray}\label{BellIneqcorr}
\mathcal{I}_3(\xi)&\!\!\!:=\!\!\! &a(\xi)\langle A_1B_1\rangle+a^{*}(\xi)\omega\langle A_1 B_2\rangle+a(\xi)\langle A_2 B_2\rangle+a^{*}(\xi)\langle A_2 B_1\rangle+\mathrm{c.c.}\nonumber\\
&\!\!\!=\!\!\!&2\mathrm{Re}\left[\langle A_1\bar{B}_1\rangle+\langle A_2\bar{B}_2\rangle\right]
\leq \mathcal{C}_d(\xi),
\end{eqnarray}
where $a(\xi)=1-\xi \omega$ 
, and $\bar{B}_1=a(\xi)B_1+a^*(\xi)\omega B_2$ and $\bar{B}_2=a(\xi)B_2+a^*(\xi)B_1$.

\subsection{Self-testing with multidimensional Bell inequalities} 

One of the applications of non-locality, intensively studied in recent years, is self-testing, which is a device-independent method to certify a given quantum state and measurements based on the observation of non-local correlations in a Bell test. In particular, the certification can be based on the violation of some Bell inequality. Let us formulate self-testing in a more detailed way, concentrating on state self-testing. Imagine that a device performs a Bell experiment on a certain unknown state $\ket{\psi'}$. The problem of state self-testing consists in deciding from the observed correlations $\{p(ab|xy)\}$ whether the state $\ket{\psi'}$ is equivalent to some reference state $\ket{\psi}$ in the sense that there exists a local isometry $\Phi=\Phi_A\otimes \Phi_B$ such that 
\begin{equation}\label{selftest}
\Phi(\ket{\psi'})=\ket{\mathrm{junk}}\otimes \ket{\psi},
\end{equation}
with $\ket{\mathrm{junk}}$ being some quantum state representing unimportant degrees of freedom. If from the correlations $\{p(ab|xy)\}$ one is able to conclude the existence of
$\Phi$ and a state $\ket{\psi}$, then we say that the state $\ket{\psi}$ has been self-tested. 

Although this problem is quite simple to formulate, it is generally difficult to solve analytically. However, there exist numerical approaches for self-testing such as the one proposed in \cite{Yang2014}. These numerical methods are particularly interesting because they provide self-testing results that are more robust to experimental imperfections than 
the analytical ones. In particular, the method of \cite{Yang2014} allows one to lower bound the fidelity between both states $\ket{\psi'}$ and $\ket{\psi}$ given the violation of some Bell inequality. 

In more precise terms, this method solves the following semidefinite program (SDP)
\begin{eqnarray}
&f = \text{min } & \langle\psi|\rho_{\text{swap}}|\psi\rangle \\ \nonumber
&\text{such that } & c \in \mathcal{Q}_{n} \\ \nonumber
&& \mathcal{I}(\xi) = \beta \\ \nonumber
&& \rho_{\text{swap}} \geq 0, \quad \text{Tr}(\rho_{\text{swap}} ) = 1 \\ \nonumber
&& \Gamma_A \geq 0, \Gamma_B \geq 0,
\end{eqnarray}
where $|\psi\rangle$ is the reference state and $\rho_{\text{swap}} $ is the state in the untrusted device on which the swap operation $\mathcal{S}$ is applied. It is given by  
\begin{equation}
\rho_{\text{swap}}  = \text{tr}_{AB} \left[ \mathcal{S} \hat{\rho}_{AB} \otimes |00\rangle\langle 00|_{A'B'} \mathcal{S}^{\dagger}\right]
\end{equation}
with $\hat{\rho}_{AB} = \proj{\psi'}$, and auxiliary qudits in $A'$ and $B'$. The swap operation $\mathcal{S} = \mathcal{S}_{AA'} \otimes \mathcal{S}_{BB'}$ is composed of local unitaries $\mathcal{S}_{AA'}$ and $\mathcal{S}_{BB'}$, so that the operation $\mathcal{S}$ together with the introduction of the auxiliary qudits constitute the local isometry evoked in (\ref{selftest}). The elements of $c$ are given by 
\begin{equation}
\left\{ c_{\{\mathbbm{1}\}} = \text{tr}(\hat{\rho}_{AB} \mathbbm{1}), c_{\{\hat{M}_{a|x}\}} = \text{tr}(\hat{\rho}_{AB} \hat{M}_{a|x}), \cdots,  c_{\{\hat{M}_{a|x} \hat{M}_{a'|x'} \hat{M}_{y|b}\}} = \text{tr}(\hat{\rho}_{AB} \hat{M}_{a|x} \hat{M}_{a'|x'} \hat{M}_{y|b}), \cdots \right\}, 
\end{equation}
and $\mathcal{Q}_{n}$ are the outer approximations of the quantum 
set obtained with level $n$ of the NPA hierarchy (when $n \rightarrow \infty$, then $\mathcal{Q}_{n} = \mathcal{Q}$). The constraint $c \in \mathcal{Q}_{n}$ expresses the (relaxed) condition that the behavior of the devices must be quantum. 
$\Gamma_A$ and $\Gamma_B$ are so-called localizing matrices (see \cite{Yang2014}), and $\beta$ is the value of the Bell expression 
$\mathcal{I}(\xi)$. When $\beta = \beta_{\max}$, we expect to obtain a minimum fidelity $f=1$, in order to conclude that the self-test is successful.

To sum up, solving this problem amounts to finding the quantum state that minimizes its fidelity to the reference state while remaining compatible with the experimentally observed correlations. Self-testing requires an almost ideal experimental setting, as by being far from the maximal quantum violation of a Bell inequality one will rapidly find orthogonal states yielding the same observed statistics, thus rendering self-testing impossible.

By using this SWAP method it was shown in \cite{Yang2014} that the partially entangled state 
(\ref{partially}) with $\gamma=(\sqrt{11}-\sqrt{3})/2$ can be self-tested from the maximal 
violation of the CGLMP Bell inequality, whereas in Ref. \cite{SATWAP} that the maximally entangled state of two qutrits can be self-tested from the violation of the Bell inequality (\ref{BellIneq1}). 

\begin{figure} [ht!]
  {\includegraphics[width=.3\linewidth]{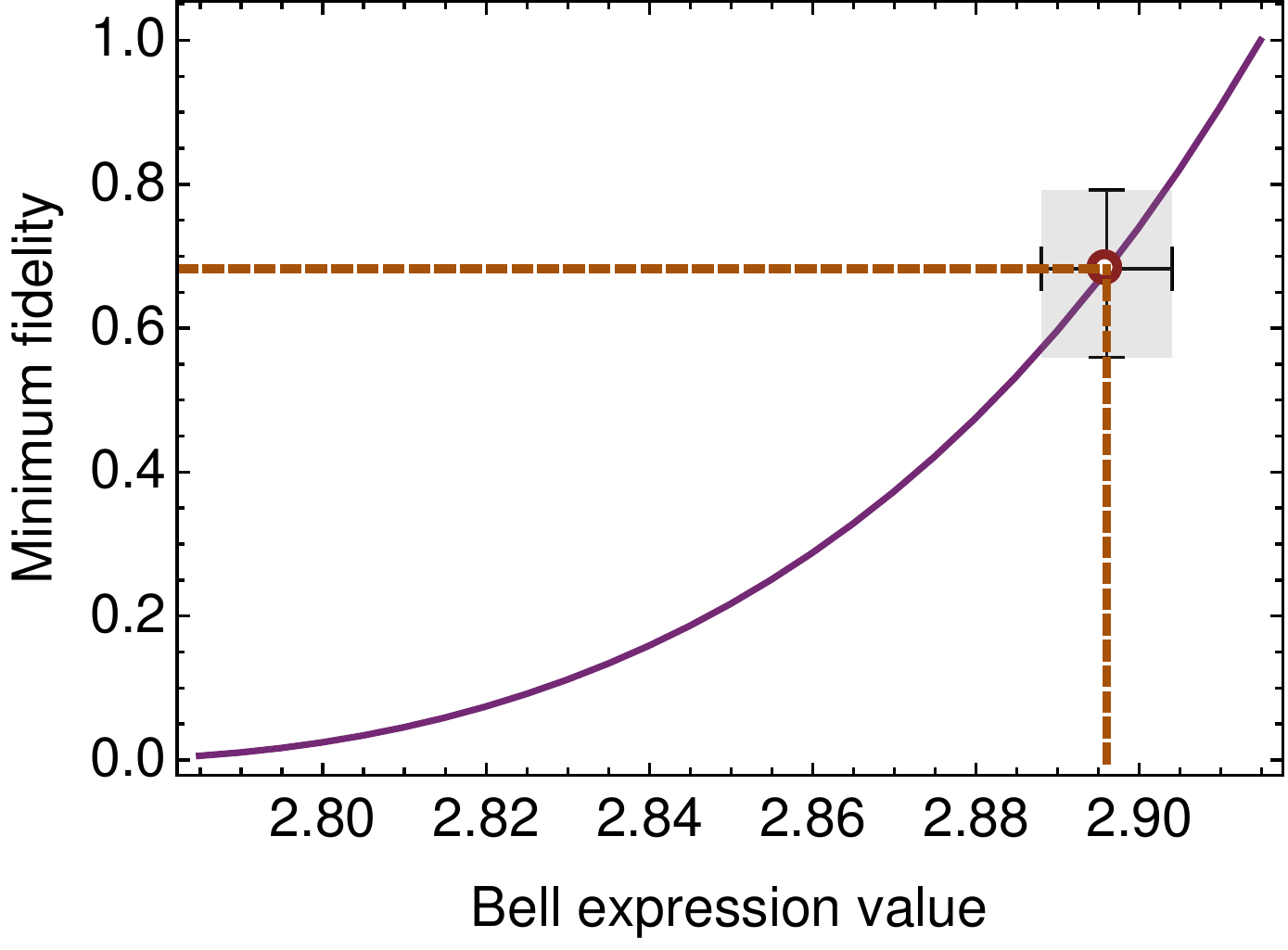}}\hfill
  {\includegraphics[width=.3\linewidth]{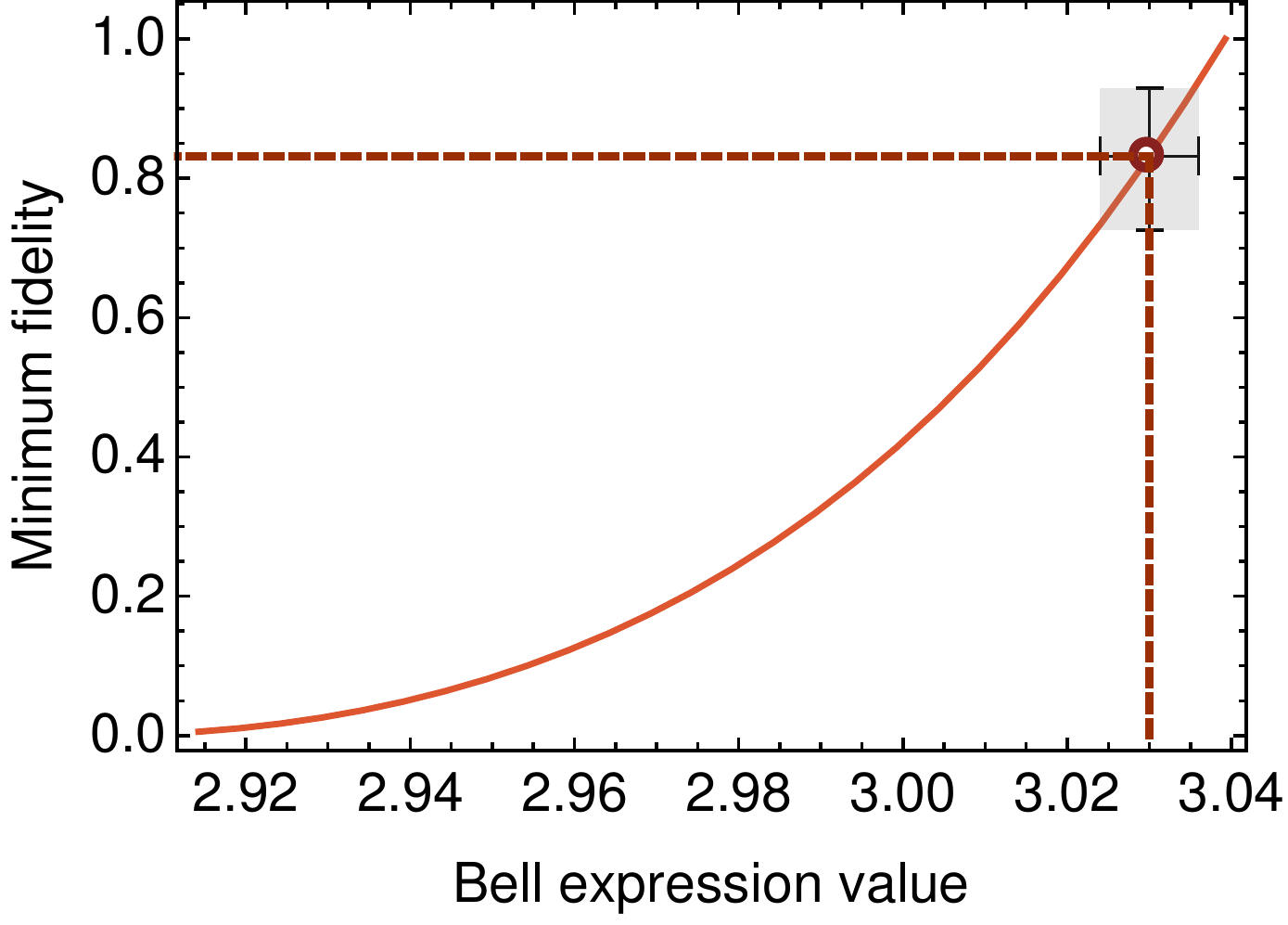}}\hfill
  {\includegraphics[width=.3\linewidth]{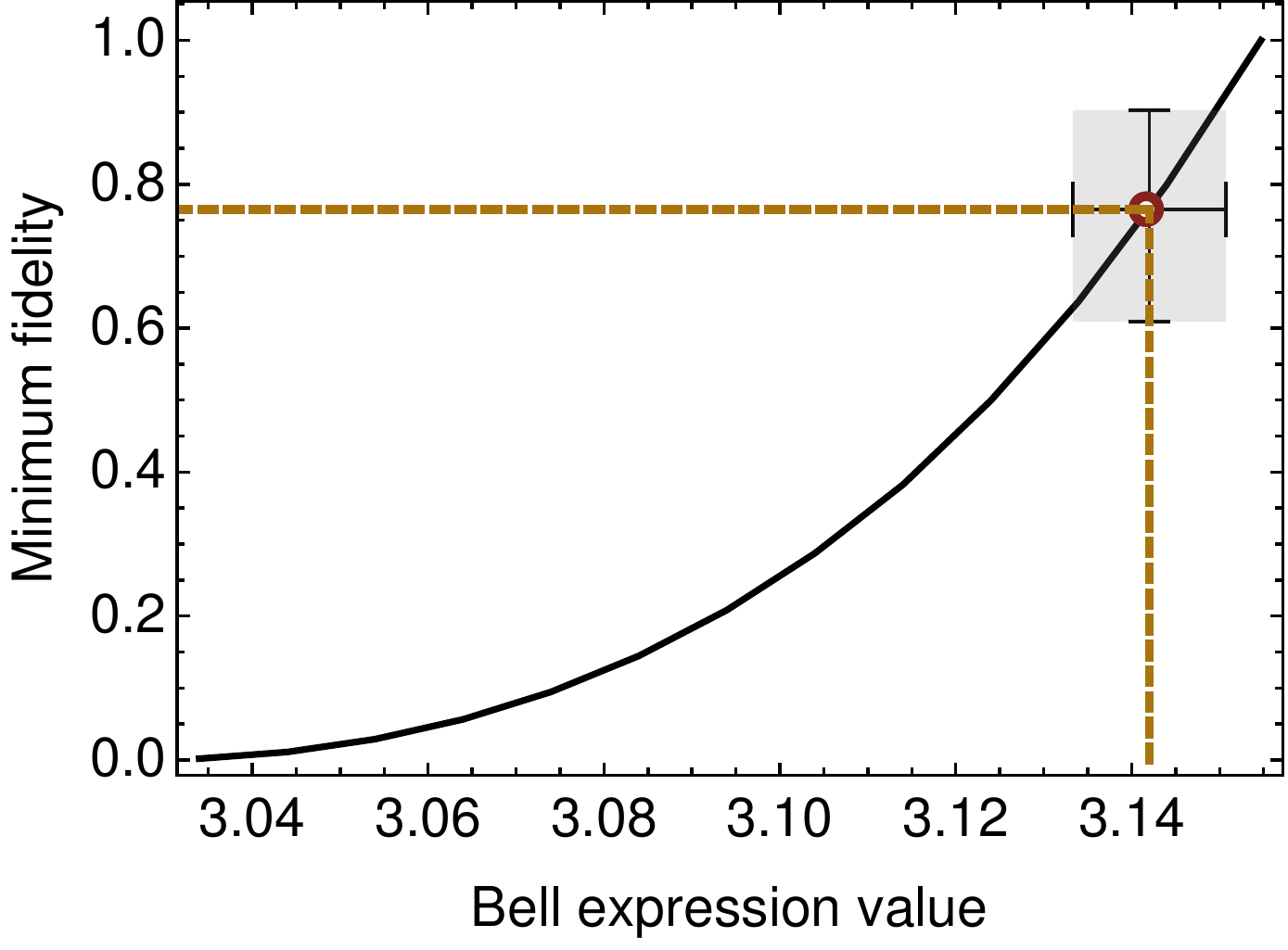}}
\caption{Minimum fidelity of state in the untrusted apparatus to the reference state $|\psi_{\gamma}\rangle$ for three values of $\gamma$, as a function of the violation of the corresponding Bell expression $I(\alpha)$. From left to right, $\gamma = 0.7923$, $\gamma = 0.9$ and $\gamma = 1$. At the maximal violation, the fidelity is equal to $1$, meaning that the quantum state used in the Bell experiment must be equal to the reference state. For lower violations, the fidelity decreases. The self-tested fidelities for the violations measured experimentally are also depicted in the figure, with error bars. 
In Fig.4C the three fidelity curves are joined, by normalizing the Bell expression value in terms of the relative quantum violation, therefore they become easily comparable. Note that, after the transformation, the three expressions become $0$ at the LHV limit and $1$ at their maximal quantum violation. 
\label{figSI_selftest}}
\end{figure}

Here we applied this method also to another state of the class (\ref{partially}), with the value $\gamma = 0.9$, which indicates that states of the form (\ref{partially}) can be self-tested by using the corresponding Bell inequalities (\ref{class}). We also computed the self-tested fidelity for experimental points obtained for $\gamma = (\sqrt{11} - \sqrt{3})/2$, $\gamma = 0.9$, and $\gamma = 1$. Complete results with robustness for values of are displayed in Figure \ref{figSI_selftest}, and in Tables \ref{table1} and \ref{table2}. Note that for the values $\gamma=1$ and $\gamma=(\sqrt{11} - \sqrt{3})/2$, which correspond to the SATWAP and CGLMP inequalities, respectively, curves as those appearing in the figures were already obtained and presented in \cite{Yang2014} and \cite{SATWAP}, except for the experimental points. 

\begin{table}[ht!]
\centering
\begin{tabular}{ c   c | c   c | c   c }
\hline \hline  \noalign{\smallskip} 
\multicolumn{2}{ c| }{$\gamma = 0.7923$}
& \multicolumn{2}{ c| }{$\gamma = 0.9$} 
& \multicolumn{2}{c}{$\gamma = 1$}  \\ \hline  \noalign{\smallskip} 
Value of $I(\xi)$ & Min. fidelity & Value of $I(\xi)$ & Min. fidelity & Value of $I(\xi)$ & Min. fidelity \\ \hline  
2.9149 & 1 & 3.0392 & 1 & 3.1547 & 1 \\  
2.9049 & 0.8193 & 3.0292 & 0.8189 & 3.1440 & 0.7992 \\  
2.8949 & 0.6638 & 3.0192 & 0.6615 & 3.1340 & 0.6372 \\  
2.8849 & 0.5319 & 3.0092 & 0.5273  & 3.1240 & 0.4993 \\  
2.8749 & 0.4207 & 2.9992 & 0.4138 & 3.1140 & 0.3823 \\  
2.8649 & 0.3276 & 2.9892 & 0.3187 & 3.1040 & 0.2871 \\  
2.8549 & 0.2500 & 2.9792 & 0.2398 & 3.0940 & 0.2082 \\  
2.8449 & 0.1858 & 2.9692 & 0.1748 & 3.0840 & 0.1446 \\  
2.8349 & 0.1334 & 2.9592 & 0.1223 & 3.0740 & 0.0945 \\  
2.8249 & 0.0914 & 2.9492 & 0.0807 & 3.0640 & 0.0564 \\  
2.8149 & 0.0585 & 2.9392 & 0.0488 & 3.0540 & 0.0289 \\ \hline
\end{tabular}
\caption{Minimum fidelity of state in the untrusted apparatus to the states $|\psi_{0.7923}\rangle$, $|\psi_{0.9}\rangle$ and $|\psi_{1}\rangle$ respectively, versus the violation of the corresponding $I(\xi)$ with $\xi = 1$, $\xi = 0.6451$, and $\xi = \frac{\sqrt{3} - 1}{2}$ respectively. At the maximal violation, the fidelity is equal to $1$, meaning that the quantum state used in the Bell experiment must be equal to the reference state. For lower violations, the fidelity decreases. \label{table_swap}}
\label{table1}
\end{table}

\begin{table}[ht!]
\centering
\begin{tabular}{ c   c | c   c | c  c }
\hline \hline \noalign{\smallskip} 
\multicolumn{2}{ c| }{$\gamma = 0.7923$}
& \multicolumn{2}{ c| }{$\gamma = 0.9$} 
& \multicolumn{2}{ c }{$\gamma = 1$}  \\ \hline \noalign{\smallskip} 
Value of $I(\xi)$ & Min. fidelity & Value of $I(\xi)$ & Min. fidelity & Value of $I(\xi)$ & Min. fidelity \\ \hline 
2.904 & 0.8051 & 3.036 &  0.9394 & 3.1507 & 0.9212 \\  
2.896 & 0.6804 & 3.030 & 0.8323 & 3.1420 & 0.7992 \\  
2.888 & 0.5711 & 3.024 & 0.7339 & 3.1333 & 0.6372 \\ \hline
\end{tabular}
\caption{Minimum fidelity of state in the untrusted apparatus to the states $|\psi_{0.7923}\rangle$, $|\psi_{0.9}\rangle$ and $|\psi_{1}\rangle$ respectively, for experimentally observed values of the violation of the corresponding $I(\xi)$ with $\xi = 1$, $\xi = 0.6451$, and $\xi = \frac{\sqrt{3} - 1}{2}$ respectively. These data are used to plot the data points and error bars in Fig. 4C of the main text, and Fig. \ref{figSI_selftest}. \label{table_swap2}}
\label{table2}
\end{table}

\section{Multidimensional steering}
In the one-sided device-independent (1SDI) scenario both the measuring devices used by Alice and the quantum state shared between Alice and Bob are uncharacterized (or untrusted). The observable data in a so-called steering test are the probability distribution of Alice's measurement outcomes, conditional on her choice of measurement $p(a|x)$, and the states prepared, or `steered' for Bob, given Alice's measurement and the shared state $\hat{\rho}_{a,x}$. In particular, assuming that the true state shared between Alice and Bob is $\hat{\rho}_d$, and that Alice's true measurements have POVM elements $\hat{M}_{a|x}$, then
\begin{align}
p(a|x) &= \text{Tr}\[(\hat{M}_{a|x}\otimes \mathbb{I}) \hat{\rho}_d\], \\
\hat{\rho}_{a|x} &= \text{Tr}_A\[(\hat{M}_{a|x}\otimes \mathbb{I}) \hat{\rho}_d\]/p(a|x).
\end{align}
This data is said to have a `local-hidden-state' (LHS) model if there exists a hidden variable $\lambda$ and collection of hidden states $\hat{\rho}_\lambda$, distributed according to $\varrho(\lambda)$, along with a response function $p(a|x,\lambda)$ that is able to reproduce it. In particular, if
\begin{align}
p(a|x) &= \int d\lambda \varrho(\lambda) p(a|x,\lambda) \\
\hat{\rho}_{a,x} &= \frac{\int d\lambda \varrho(\lambda) p(a|x,\lambda) \hat{\rho}_\lambda}{\int d\lambda \varrho(\lambda) p(a|x,\lambda)}
\end{align}
Such a model provides a classical explanation of the steered states: all correlations between Alice's outcomes and Bob's states are mediated by the hidden variable $\lambda$, which controls Alice's measurements through the response function $p(a|x,\lambda)$, and Bob's state through $\hat{\rho}_\lambda$. 

LHS models are not able to reproduce all the predictions of quantum theory. In particular, it can be shown that all LHS models satisfy linear constraints, known as EPR steering inequalities, of the form
\begin{equation}
\beta_d = \sum_{\substack{a=b \\ x=y}} p(a|x)\text{Tr}\[\hat{M}_{b|y}\hat{\rho}_{a|x}\] \leq \beta_{lhs}
\end{equation}
which can be violated by data $p(a|x)$, $\hat{\rho}_{a|x}$ arising in quantum theory. Violations of EPR steering inequalities certify, in a one-sided device-independent manner, that the underlying state shared by Alice and Bob is entangled. Experimental values obtained for $\beta_d$, up to local dimension $15$, are shown in Fig.~5A in the main text, and summarized in Table~\ref{tableSteer}.

\begin{table}[htb]
\centering
\begin{tabular}{c| c c c c c c c c } 
 \hline \hline \noalign{\smallskip} 
Dim & 2 & 4 & 6 & 8 & 10 & 12 & 14 & 15 \\
\hline \noalign{\smallskip}
$\beta_{lhs}$  & $1.707$ & $1.5$ & $1.408 $ & $1.354$  & $1.316$  & $1.289$  & $1.267$ & $1.258$ \\
$\beta_d$  & $1.990\pm 0.001$ & $1.944 \pm 0.001$ & $1.901 \pm 0.002 $ & $1.849 \pm 0.002 $  & $1.822 \pm 0.002 $  & $1.799 \pm 0.003 $  & $1.768 \pm 0.002 $ & $1.70\pm 0.003$
\\ \hline 
\end{tabular}
\caption{ Experimental values $\beta_d$ for multidimensional steering correlations. The steering is certified by violating the steering inequality $\beta_d\leq \beta_{lhs}$, where $\beta_{lhs}=1+1/\sqrt[]{d}$ is the classical bound for LHS models. Violating the LHS bound demonstrates the presence of entanglement and steering. These data are used to plot Fig~5A in main text. }
\label{tableSteer}
\end{table}

\section{Device-independent and one-sided device-indepedent randomness generation}
Violations of Bell and EPR Steering inequalities can be used to generate, in a device-independent or one-sided device-independent manner, respectively, private randomness. The intuition is that, since a bipartite entangled state is prepared, the reduced state of each part is necessarily mixed, from which it follows that measurement outcomes are fundamentally uncertain. Moreover, due to monogamy of entanglement, an Eavesdropper cannot be perfectly correlated with these measurement outcomes, and hence there must be some residual uncertainty for them, even if they have additional  side information about the state and measurements. This intuition can be formalized, and it can be shown that the guessing probability (i.e. the predictability) of the outcomes of the measurements of the devices, can be bounded.

More precisely, we will assume that a party who wants to generate private randomness has a source of entangled particles, and two independent measuring devices in their possession, which will be used to measure each subsystem. We will consider two scenarios: (i) the (fully) DI scenario, where neither the source of entanglement nor either measuring device is characterised or trusted. Here, entanglement will be certified using the violation of Bell inequalities. Global randomness will be generated from the pair of outcome strings of both measuring devices. (ii) the one-sided DI scenario, where one of the two measuring devices  is assumed to be characterised. Here entanglement will be certified using the violation of EPR Steering inequalities. Local randomness will now only be generated from the outcome string of the single uncharacterised measuring device. In each case, the following protocol will be followed: (I) Using a source of (public or private) randomness, the party will choose the settings $(\mathbf{x},\mathbf{y})$ or $\mathbf{x}$  to perform $n$ rounds of a Bell or steering test, respectively. (II) The party will perform the chosen measurements for the SATWAP Bell inequality, or the Steering inequality \eqref{eq:steering} on $n$ successive pairs of particles, and record the outcome strings $\mathbf{a}$ and $\mathbf{b}$. (III) An estimate $\tilde{I}_d^{est}$ or $\beta_d^{est}$ for the violation $\tilde{I}_d$ or $\beta_d$ will be calculated. This will then be used in the following programs to find DI or 1SDI bounds on the predictability of either (i) each pair of symbols $(a_i,b_i)$ of the strings $(\mathbf{a},\mathbf{b})$ in the DI case (ii) each symbol $a_i$ of the string $\mathbf{a}$ in the 1SDI case. The programs are as follows \cite{Pironio:2010,}. For the DI case we have:
\begin{align}
g^{DI}(\tilde{I}_d^{est}) := \max_{x,y} \max_{\{p(a b,e|xy)\}} &\quad \sum_{e = (a,b)} p(a b,e|xy) \nonumber \\
s.t & \quad \tilde{I}_d  = \tilde{I}_d^{est} \\
& \quad p(a b,e|xy) \in Q
\end{align} 
where the optimisation variables are the collection of probability distributions $\{p(a b,e|x,y)\}$ for the two partners in the protocol, and that of the Eavesdropper (who has the additional side information $e$, which without loss of generality can be taken to be a guess for the pair $(a,b)$); $\tilde{I}_d$ is the value obtained for the SATWAP inequality that arises from $p(a b|x y) := \sum_e p(a b, e|xy)$, i.e. from the marginal distribution of the devices used; $Q$ represents the set of quantum correlations. This last constraint cannot be directly enforced in general, but can be relaxed to $p(a b,e|xy) \in Q_k$, where $Q_k$ stands for level k of the NPA hierarchy \cite{npa2007}. For each pair of values $(x,y)$ the inner maximisation is thus approximated by an SDP. By solving it for all pairs, and taking the worst case predictability, one can obtain a bound on how well Eve can guess each pair of output symbols of the two devices. 

For the 1SDI case, the corresponding program is \cite{CavSkr17InPrep}

\begin{align}\label{e:SDP guessing prob}
g^{1SDI}(\beta_d^{est}) := \max_{x} \max_{p(a,e|x),\hat{\rho}_{a,e|x}} & \quad \sum_{a=e} p(a,e|x) \nonumber \\
\text{s.t}&\quad \sum_{\substack{a=b\\e \\ x=y}} p(a,e|x)\text{Tr}\[\hat{M}_{b|y}\hat{\rho}_{a,e|x}\] = \beta_d^{est}, \nonumber \\
&\quad \sum_a p(a,e|x)\hat{\rho}_{a,e|x} = \sum_a p(a,e|x')\hat{\rho}_{a,e|x'} \quad \forall e,x,x' \\
& \quad \hat{\rho}_{a,e|x} \geq 0, \quad \text{Tr}\[\hat{\rho}_{a,e|x}\] = 1, \quad \forall a,e,x,\nonumber \\
&\quad p(a,e|x) \geq 0, \quad \forall a,e,x, \quad \sum_{a,e} p(a,e|x) = 1,\quad \forall x. \nonumber
\end{align}
Here the optimization variables are the joint probability distribution $p(a,e|x)$ of the outcomes $a$ of the first  measuring device and $e$, (which is now Eve's guess of $a$), and the states prepared for the second device (by Eve and Alice's measurement) $\hat{\rho}_{a,e|x}$. The first condition, similar to above, is the constraint that Alice and Bob observe a given violation of the steering inequality \eqref{eq:steering}, while the remaining conditions ensure that this is a valid quantum strategy of Eve (i.e. that it obeys the no-signaling condition, and that Eve uses a valid quantum state and valid probability distribution. It can be shown that by a change of variables (to $\hat{\sigma}_{a,e|x} := p(a,e|x)\hat{\rho}_{a,e|x}$) that the inner constrained optimization problem is an SDP. Hence, as above, by solving the SDP for each value of $x$, and taking the worst case, a bound on the predictability of the string $\mathbf{a}$ is obtained. 

The above programs give us the desired bounds on the predictability for Eve for the outcomes,
\begin{align}
P_g(AB) &\leq g^{DI}(\tilde{I}_d^{est}), \\
P_g(A) &\leq g^{1SDI}(\beta_d^{est}).
\end{align}
Associated to these DI and 1SDI upper bounds on the guessing probability of Eve, we obtain the corresponding lower bounds on the min-entropy (amount of randomness) that can be extracted by the user per output symbol
\begin{align}
H_{min}(AB|E) \geq f^{DI}(\tilde{I}_d^{est}) &:= -\log_2 g^{DI}(\tilde{I}_d^{est}) \\
H_{min}(A|E) \geq f^{1SDI}(\beta_d^{est}) &:= -\log_2 g^{1SDI}(\beta_d^{est})
\end{align}
(IV) Finally, the user applies an appropriately chosen randomness extractor to the pair of output strings $(\mathbf{a},\mathbf{b})$ in the DI case, or the output string $\mathbf{a}$ in the 1SDI case (along with a small random seed), which will produce $n f^{DI}(\tilde{I}_d^{est})$ or $nf^{1SDI}(\beta_d^{est})$ bits of randomness, that with overwhelming probability will be uniformly random and private from the Eavesdropper. The experimentally obtained values for the bounds $f^{DI}$ and $f^{1SDI}$, obtained preparing maximally entangled states, are reported in Fig.~5 in the main text. The experimental results for all the states tested, including partially entangled states, are summarized in Table~\ref{tableRandomSemiDI} and Table~\ref{tableRandomDI}.

\begin{table}[ht!]
\centering
\begin{tabular}{c | c | c| c  }
\hline \hline \noalign{\smallskip}
\multicolumn{1}{ c }{}
& \multicolumn{3}{ c }{One-sided DI local randomness} \\ \hline \noalign{\smallskip}
Dim & $f^{1SDI}$ lower value & $f^{1SDI}$ mean value & $f^{1SDI}$ higher value \\ \hline  \noalign{\smallskip}
2 & 0.7277 & 0.7438 & 0.7614  \\  
4 &  1.0611 & 1.0676 & 1.0742  \\  
6 & 1.1247 & 1.1352 & 1.1458  \\  
8 &  1.0653 & 1.0738 & 1.0823  \\  
10 & 1.0755 & 1.0834 & 1.0914  \\ 
12 & 1.0745 & 1.0859 & 1.0974  \\ 
14 & 1.0289 & 1.0359 & 1.0430  \\  
15 & 0.8356 & 0.8430 & 0.8504  \\ \hline
\end{tabular}
\caption{Certified local randomness in the one-sided DI scenario by the violation of steering inequalities. 
Reported are the mean number of global bits of randomness $f^{1SDI}$ obtained from the observed violations of the steering inequalities, and correspondent lower (higher) values correspondent to -($+$)1$\sigma$ confidence interval for the steering value, calculated using Poissonian photon statistics.. These data are used to plot Fig. 5B in main text. }
\label{tableRandomSemiDI}
\end{table} 

\begin{table}[ht!]
\centering
\begin{tabular}{ c | c |c| c  }
\hline \hline \noalign{\smallskip}
\multicolumn{1}{ c}{}
& \multicolumn{3}{ c }{Fully-DI global randomness}  \\ \hline  \noalign{\smallskip}
Dim & $f^{DI}$ lower value & $f^{DI}$ mean value & $f^{DI}$ higher value \\ \hline   \noalign{\smallskip}
2 &  0.8903 & 0.9687 & 1.0992  \\  
3 & 1.3308 & 1.4412 & 1.6130  \\ 
4 & 1.5922 & 1.8227 & 2.1400  \\  
5 & 0.9500 & 1.0036 & 1.0606  \\  
6 & 0.3085 & 0.3284 & 0.3490  \\  
7 & 0.2835 & 0.3001 & 0.3172  \\  
8 & 0.4417 & 0.4707 & 0.5008  \\ \hline \noalign{\smallskip}
$d =3, \gamma=0.9$ & 1.4566 & 1.5415 & 1.6693 \\ 
$d =3, \gamma=0.7923$ & 1.3981 & 1.4708 & 1.5651 \\ \hline
\end{tabular}
\caption{Certified global randomness in the fully DI scenario by the violation of Bell inequalities. 
Reported are the mean number of global bits of randomness $f^{DI}$ obtained from the observed violations of the SATWAP inequalities, and correspondent lower (higher) values correspondent to -($+$)1$\sigma$ confidence interval for the Bell value, calculated using Poissonian photon statistics. These data are used to plot Fig. 5C in main text. 
For the partially entangled qutrit states $(\ket{00}+\gamma\ket{11}+\ket{22})/(\sqrt{2+\gamma^2})$, global DI randomness are verified by the violation of Bell inequalities as for the maximally-entangled cases.  
$\gamma=1$ represents the maximally entangled state, while $\gamma=(\sqrt{11}-\sqrt{3})/2$ represents the partially entangled state that maximally violates CGLMP inequalities. 
}
\label{tableRandomDI}
\end{table}

\section{On-chip multidimensional quantum key distribution} 
\begin{figure}[h]
\centering
\includegraphics[width=0.72\textwidth]{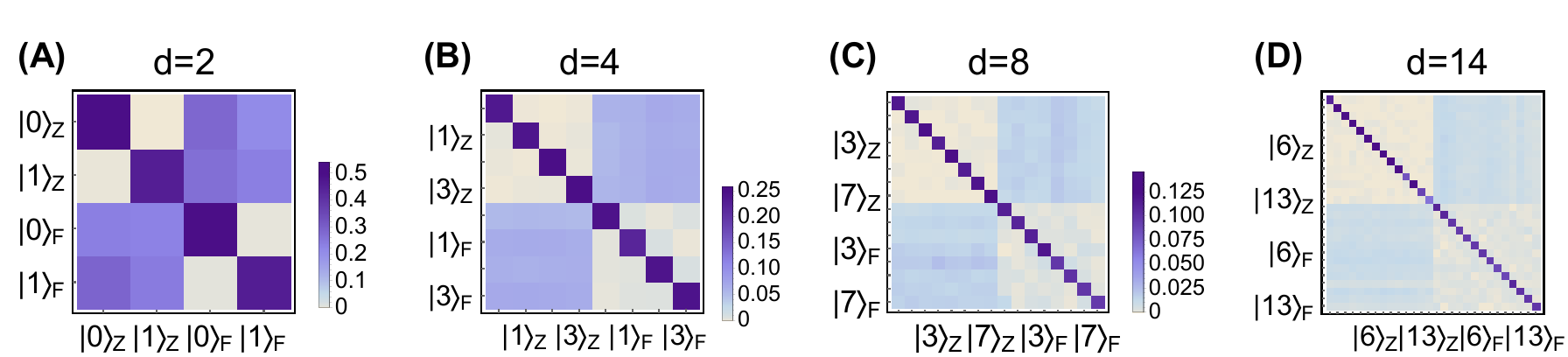}
\caption{Correlations matrices for multidimensional entanglement based QKD. 
Alice and Bob are on the single chip and measured in the mutually unbiased bases.  
The label $Z$ indicates computational basis and $F$ to the Fourier basis. 
}  
\label{fig:T_QKD}
\end{figure}

Quantum key distribution (QKD) allows two remote parties to establish an unconditionally secure key by using a quantum channel in combination with a public channel (classical channel), and represents one of the most renowned applications of quantum technologies~\cite{Scarani2009,PhysRevLett.88.127902}.
While most implementations have focused on qubits, an increasing number of multidimensional QKD demonstrations have been reported in recent years. In our experiment we use a generalized entanglement-based version of the BB84 protocol for multi-dimensions, initially proposed and analyzed in Ref.~\cite{PhysRevLett.88.127902}., and test it for different dimensions ($d=2,4,8,14$). In Fig.~\ref{fig:T_QKD} we report experimental correlations data for multidimensional mutually unbiased bases required for QKD (computational and Fourier bases)~\cite{Scarani2009,PhysRevLett.88.127902}, obtaining fidelities $99.78 \pm 0.02 \%$, $99.24 \pm 0.02 \%$, $97.70 \pm 0.03 \%$, and $96.27 \pm 0.04 \%$  for local dimensions $2$, $4$, $8$ and $14$, respectively.  
For the security analysis we use the approaches reported in Refs.~\cite{Scarani2009,PhysRevLett.88.127902}. The final secret key rate (per coincidence) can be obtained as
\begin{equation}
R_{sk} \geq max (I_{AB}-I_{AE},I_{AB}-I_{BE}),
\end{equation}
where $I_{AB}$ represents the mutual information between Alice and Bob, given as a function of the fildelity $F$ and the dimension $d$ by $I_{AB}(F,d)= log_2(d) + F \, log_2 (F) + (1-F) \, log_2 (\frac{1-F}{d-1})$. Similarly, $I_{AE}$ or $I_{BE}$ stands for the mutual information between Alice and Eve or Bob and Eve, and represent the amount of information an Eavesdropper can get about the key obtained by the two partners. This value depends on the type of attacks that Eve is allowed to perform. In Table~\ref{t:QKD} we report the maximal values of the quantum bit error rate (QBER), given by the infidelity $\text{QBER}=1-F$, required to achieve a positive key rate in the cases where Eve is allowed coherent attacks ($\text{QBER}_{(\text{Th})}^{\text{Coh}}$) or individual attacks ($\text{QBER}_{(\text{Ind})}^{\text{Coh}}$), which were analytically obtained in Ref.~\cite{PhysRevLett.88.127902}, together with the experimentally measured values of $F$ and of the QBER.  
It can be observed that in general higher dimensionality corresponds to higher tolerance to noise and higher photon information efficiency. The experimentally measured QBER values are considerably within the required bounds, and the associated lower bounds on the secure key-rates, quantified as bits per coincidence, are also reported in Table~\ref{t:QKD} for different dimensions. To be noted that in this implementation we assume the reconciliation efficiency parameter to be equal to identity and the fraction of multi-pair contribution consists of independent uncorrelated pairs. The key rates increase with the dimensionality of the system, experimentally demonstrating that increasing the dimension of the system the information capacity can be significantly improved. \\

\begin{table}[htb]
\centering
\begin{tabular}{c| c c c c c} 
 \hline \hline \noalign{\smallskip} 
Dim & Fid. (\%) &$\text{QBER}_{(\text{Exp})}$ (\%)  &$\text{QBER}_{(\text{Th})}^{\text{Coh}} (\%)$ & $\text{QBER}_{(\text{Th})}^{\text{Ind}} (\%)$ & $R_{sk}$ [bpc]\\
\hline \noalign{\smallskip}
$2$ & $99.78\pm 0.02$ &$0.22 \pm  0.02$ & $\leq 11.00$& $\leq 14.64$ &(0.5) $0.477 \pm 0.002 $ \\
$4$ &$99.24\pm 0.02$ &$0.76 \pm 0.02$ & $\leq 18.93$& $\leq 25.00$ &(1.0) $0.924 \pm 0.002$ \\
$8$ &$97.70\pm 0.03$ &$2.30 \pm 0.03$ &$\leq 24.70$& $\leq 32.32$ &(1.5) $1.277 \pm 0.003$ \\
$14$ &$96.27\pm 0.04$ &$3.73 \pm 0.03$ & $\leq 28.24$& $\leq 36.64$ &(1.903) $1.536 \pm 0.003$ \\ \hline 
\end{tabular}
\caption{\label{t:QKD} {Security analysis for device-dependent QKD (BB84-type) in different dimensions.} Reported are the experimentally measured fidelities and associated experimental QBER and secure key rate ($R_{sk}$, in bits per coincidences) values for dimension up to 14, together with the theoretical maximal QBER bounds for coherent ($\text{QBER}_{(\text{Th})}^{\text{Coh}}$) and individual ($\text{QBER}_{(\text{Th})}^{\text{Ind}}$) attacks. For $R_{sk}$, quantities in brackets indicate ideal values obtained with perfect correlations, to be compared with the experimental data. The analysis is performed following the approach developed in Ref.~\cite{PhysRevLett.88.127902}. }
\end{table}

While the previous analysis was performed in a device-dependent scenario, the measured non-local correlations allows us to perform also a security analysis in device-independent quantum key distribution settings~\cite{SATWAP} (under the fair-sampling assumption). Here the bound on the key rate is given by
\begin{equation}
R^{DI}_{sk}\geq H_{min} (A|E) - H(A|B),
\end{equation}
where, equivalently as described in the randomness generation sections, $H_{min}(A|E)$ represents the min-entropy of Alice's outcomes for Eve, and $H(A|B)$ is the conditional Shannon entropy between Alice's and Bob's outcomes distribution, in the correlated bases used to construct the key. $H_{min}(A|E)$ can be bounded numerically, based on the measured violation of the SATWAP inequality, using the NPA hierarchy as before~\cite{npa2007}. The experimental secure key rates for various dimensions for the DI scenario are reported in Table~\ref{tableDIQKD}. Positive DI key rates are observed for dimensions up to $d=8$. For $d=4$ a rate higher than 1 bit per symbol is obtained.
\ \\

We remark that in our particular implementation the generation and the measurement processes are on the same integrated photonic structure, meaning that Alice, Bob and the entanglement source are at a mm-scale distance. However, recent developments in chip-to-chip distribution of multidimensional path-encoded photonic states can be adopted to distribute the entanglement generated from our chip~\cite{Wang2016,Ding:QKD}. What we show here is that our integrated scheme has the potential to be used as a source of high-dimensional entangled states for entanglement-based multidimensional quantum applications and device-independent QKD protocols.

\begin{table}[ht!]
\centering
\begin{tabular}{ c | c |c| c}
\hline \hline \noalign{\smallskip}
Dim & $R^{DI}_{sk}$ lower value & $R^{DI}_{sk}$ mean value & $R^{DI}_{sk}$ higher value\\ \hline  
\noalign{\smallskip}
2 &  0.70 & 0.76 & 0.86 \\  
4 & 1.33 & 1.53 & 1.94  \\ 
8 & 0.24 & 0.27 & 0.29 \\ 
\hline
\end{tabular}
\caption{{Security analysis for DI QKD in different dimensions.}
Reported are the mean value of the secure key rate $R^{DI}_{sk}$ obtained from the observed violations of the SATWAP inequalities, and correspondent lower (higher) values correspondent to -($+$)1$\sigma$ confidence interval for the Bell value, calculated using Poissonian photon statistics.
}
\label{tableDIQKD}
\end{table}

\ \\
\begin{algorithm}[H]
	\hrulefill
	\vskip-8pt
	\caption{Elimination on a single Mach-Zehnder}
	\vskip-7pt
	\hrulefill
	\label{algorithmPhases}
	\begin{algorithmic}
		\smallskip
		\STATE  \textbf{Input:} $\{ a_k \}$, set of complex amplitudes.
        \STATE  \textbf{Input:} $k_{PS}$, mode with the phase-shifter inside the MZI. 
        \STATE  \textbf{Input:} $k_{NoPS}$, mode without the phase-shifter.
        \STATE  \textbf{Input:} $k_\times$, mode whose associated amplitude is to be eliminated.
        
        \\
		
		\STATE \textbf{function} {\large \tt{MZI\_ELIMINATION}} ($\{ a_k \},k_{PS},k_{NoPS},k_\times$) :
        
        \STATE \indent \textbf{if}
		$k_\times = k_{PS}$:
		\STATE \indent\indent
 $\theta = 2 \tan^{-1}(|a_{k_{NoPS}}/a_{k_{PS}}|) $   \indent $\triangleright$ calculate phase inside MZI.
        \STATE \indent\indent
		 $\phi = \text{Arg}(a_{k_{NoPS}})-\text{Arg}(a_{k_{PS}})$   \indent $\triangleright$ calculate the required phase offset between the input modes.
         
 		\STATE\indent \textbf{else}:     
        \STATE \indent\indent
 $\theta = 2 \tan^{-1}(|a_{k_{PS}}/a_{k_{NoPS}}|) $   \indent $\triangleright$ calculate phase inside MZI.
        \STATE \indent\indent
		 $\phi = \text{Arg}(a_{k_{NoPS}})-\text{Arg}(a_{k_{PS}})+\pi$   \indent $\triangleright$ calculate the required phase offset between the input modes.        		
             
		\STATE\indent \textbf{end if}\\
        
\STATE \indent $\begin{pmatrix}
  a_{k_{NoPS}} \\
  a_{k_{PS}} \\
 \end{pmatrix} \leftarrow  \hat{M}_{MZI}(\theta)  \hat{M}_{PS}(\phi)  \begin{pmatrix}
  a_{k_{NoPS}} \\
  a_{k_{PS}} \\
 \end{pmatrix}$   \indent $\triangleright$ update modes amplitudes using the obtained phases $\theta$ and $\phi$. \\

		\STATE\indent \textbf{return} ($\theta,\phi,\{a_k\}$)\\    
        
  		\STATE \textbf{end function}\\        
    
  		\STATE \textbf{Output:} The phase $\theta$ of the phase-shifter inside the interferometer and the phase shift between the two input mode $\phi$ required to eliminate the element such that full destructive interference is obtained on the mode $k_\times$ (i.e. $a_{k_\times}$=0). The updated values of the state complex amplitudes $\{a_k\}$. \\       
		
	\end{algorithmic}
\end{algorithm} \ \\

\begin{algorithm}[H]
	\hrulefill
	\vskip-8pt
	\caption{Calculate phases for arbitrary qudit projective measurements using the triangular scheme}
	\vskip-7pt
	\hrulefill
	\label{FullalgorithmPhases}
	\begin{algorithmic}
		\smallskip
        
	\STATE  \textbf{Input:} $\{ c_k \}$, set of complex amplitudes describing the $d$-dimensional projective state $|\psi\rangle=\sum_{k=0}^{d-1} c_k |k\rangle$, where $d=2^N$.   \\
		
		\STATE \textbf{function} {\large \tt{GETPHASES}} ($\{ c_k \}$) :
		\STATE\indent Initialise $\{a_k\}=\{c_k\}$
        \STATE\indent Initialise $\{\phi_\ell\}$ as $\phi_\ell=0 \quad \forall \ell\in\{1,d\}$\\

 		\STATE\indent \textbf{for} $n=1 \to N$\indent 		
        \STATE\indent\indent \textbf{for} $i=2^{N-n} \to 1$\indent 		
        
        \STATE\indent\indent\indent $k_{bottom}=2i-1$
        \STATE\indent\indent\indent $k_{top}=2i$\\
        
         \STATE\indent\indent\indent \textbf{if} $i$ odd: \indent $\triangleright$ at each step select whether the mode whose amplitude we want to eliminate is the top or the bottom one.
         \STATE\indent\indent\indent\indent $k_\times=k_{bottom}$    
          \STATE\indent\indent\indent \textbf{else}:
         \STATE\indent\indent\indent\indent $k_\times=k_{top}$   
         \STATE\indent\indent\indent \textbf{end if} \\
         
		\STATE\indent\indent\indent \textbf{if} the phase-shifter in the $i$-th MZI of the $n$-th layer is on the bottom mode:
    
          \STATE\indent\indent\indent\indent $(\theta_{n,i},\phi,\{a_k\}) \leftarrow \text{\large \tt{MZI\_ELIMINATION}} (\{ a_k \},k_{bottom},k_{top},k_\times)$ \indent $\triangleright$ eliminate the mode and update phases.
          \STATE\indent\indent\indent\indent \textbf{foreach} $\ell \in \{(i-1)2^n+1,\ldots,(i-1)2^n+2^{n-1}\}$  \textbf{do}  $\phi_\ell \leftarrow  \phi_\ell + \phi$
		\STATE\indent\indent\indent \textbf{else}:          
         \STATE\indent\indent\indent\indent $(\theta_{n,i},\phi,\{a_k\}) \leftarrow \text{\large \tt{MZI\_ELIMINATION}} (\{ a_k \},k_{top},k_{bottom},k_\times)$ \indent $\triangleright$ eliminate the mode and update phases.
          \STATE\indent\indent\indent\indent \textbf{foreach} $\ell \in \{2^ni-2^{n-1},\ldots,2^ni\}$  \textbf{do}  $\phi_\ell \leftarrow  \phi_\ell + \phi$

          \STATE\indent\indent\indent \textbf{end if} \\          
         \STATE\indent\indent\indent  Remove the $k_\times$-th element ($ a_{k_\times} =0$) from the list $\{a_k\}$.  \indent $\triangleright$ update the list of non-zero mode amplitudes. \\      
        
         \STATE\indent\indent \textbf{end for}

         \STATE\indent \textbf{end for}
		
		\STATE \indent \textbf{return} $(\{\theta_{n,i}\}, \{ \phi_\ell\})$.\\
		
		\STATE \textbf{end function} \\

		\STATE  \textbf{Output:} The list of phases to set in the MZIs $\{\theta_{n,i}\}$, where $\theta_{n,i}$ represents the phase to be set in the $i$-th MZI of the $n$-th layer. The list of phases $\{ \phi_\ell\}$ to set on the single modes before the triangular structure.

	\end{algorithmic}
\end{algorithm}

\end{document}